\documentclass[12pt,preprint]{aastex}

\newcommand{\msun}{\ensuremath{M_{\odot}}}

\slugcomment{}

\shorttitle{Active galaxies in the SDSS}
\shortauthors{Reviglio, Helfand}

\begin{document}

\title{Active Galaxies in the Sloan Digital Sky Survey I: The Spectroscopically Unremarkable Population of the Local Universe}

\author{Pietro M. Reviglio \&  David J. Helfand}
\affil{Astronomy Department, Columbia University, New York, NY 10027}

\email{reviglio@astro.columbia.edu,djh@astro.columbia.edu}

\begin{abstract}

In a series of papers based on the FIRST and Sloan Digital Sky Surveys (SDSS), we investigate the local population of star-forming galaxies and Active Galactic Nuclei (AGN) in order to clarify the link between these two types of activity, to explore the dependence of their features on the properties of their host galaxies, and to examine the role of the environment in triggering activity. In this first paper, we present the multiwavelength database created for $\sim 150,000$ SDSS galaxies with $14.5 \le R \le 17.5$.  We compare different methods of classification for AGN activity and  show that pollution of nuclear spectra by host galaxy light leads to a serious misclassification of weak-lined AGN with increasing redshift. We develop an algorithm to correct for this bias and show that, for a fixed host luminosity, the misclassification preferentially affects redder systems, suggesting that lines in these systems are weaker. After correction for dilution, the sample displays the same fraction of AGN as in the sensitive Palomar Survey of nearby galaxies (Ho et al.). We demonstrate  that spectroscopically truly passive systems show signatures of X-ray or radio AGN activity; by stacking radio images we  establish the median nuclear radio luminosity of this class. While we confirm that the fraction of AGN with strong [OIII] emission decreases in denser environments, we show that, including the faint population of AGN, the fractional abundance of AGN \emph{increases} with increasing density, suggesting that emission-line AGN activity in denser environments is more frequent but less intense.

\end{abstract}


\section{Introduction}
\label{sec:intro}

The prevalence of activity in galaxies and its evolution through cosmic time is one of the main unsolved problems of galaxy evolution. It is now firmly established that activity in galaxies evolves: more powerful radio sources are found at higher redshifts \cite[]{Longair1966, Dunlop1990}, the star-formation rates of galaxies increase with redshift \cite[]{Lilly1996, Madau1998}, and the population of blue galaxies in more distant clusters increases with look-back time \cite[]{Butcher1978}, while the number counts of X-ray-emitting active galaxies increase with redshift in a similar way \cite[]{Hasinger1993}. Along with this evolution, it appears that galaxy morphologies have changed over time, with higher redshift galaxies appearing more disturbed and the fraction of disky galaxies increasing at higher redshifts.

The link between activity and galaxy morphology is not understood: the standard picture holds that disk galaxies are often star-forming, while the hosts of AGN are typically red bulgy systems with little ongoing star formation. Powerful radio and X-ray AGN have been found in ellipticals with old stellar populations that display no recent star-formation \cite[]{Reviglio2006, Nandra2007}. However, some evidence exists for recent star formation in both the brightest active galactic nuclei (Kauffmann et al. 2003) and the much less luminous AGN population \cite[]{deVries2007} 

In a series of papers based on the Sloan Digital Sky Survey Second Data Release  (SDSS-DR2), we analyze a local population of star-forming galaxies and active galactic nuclei. We present the sample we have constructed here in Paper I and examine the selection systematics for star-forming galaxies and AGN, focusing in particular on the differences between alternative methods of classification and the effects such differences produce on the inferred demographics of these populations in the local universe. In two subsequent papers we will present a comparative analysis of the sample's radio and spectroscopic features, and investigate the dependence of these on the physical properties of host galaxies, such as supermassive black hole mass, spin, and accretion rate, as well as on the environment in which they reside.

A clean and unbiased analysis of the local population of star-forming galaxies and AGN is crucial for understanding both the properties of different types of activity -- e.g., the onset of star-formation on galactic scales, the formation of broad and narrow emission-lines in AGN and quasars, the powering of radio jets and X-ray AGN -- and the role of environment in triggering or suppressing these types of activity.  Moreover, a sound knowledge of the local population of active galaxies is required in order to compare their properties with the high-redshift samples now becoming available. The wealth of information provided by the Sloan Digital Sky Survey (SDSS) allows an analysis of the local population of active galaxies and their environments in unprecedented detail. Nonetheless, as pointed out by Ho (2004), unbiased selection of active galaxies is not simple to achieve, especially when exploring relatively faint optical populations such as those sampled by the SDSS. 

There are many sources of bias in the spectroscopic characterization of active galaxies. Aperture bias, for example, has been shown to effect severely the interpretation of spectroscopy of star-forming galaxies, altering their inferred star-formation rates, metallicities, and reddenings \cite[]{Kewley2005}. Hopkins et al. (2003) have estimated the magnitude of aperture bias on star-formation indicators in the SDSS and have proposed an empirical correction for its effects.

Likewise, proper starlight subtraction in the spectra of AGN is particularly challenging to achieve and may result in significant incompleteness in the AGN population.  The effect of host starlight dilution of Seyfert spectra has been studied by Moran et al. (2002) who showed that for Seyfert galaxies with observed sizes comparable to the size of the spectral aperture (slit width or fiber aperture), the lines are diluted by the stellar continuum and are no longer detectable, with the consequence that the host appears as an unremarkable, passive galaxy. 

In order to avoid this problem, more sophisticated methods of stellar light subtraction have been developed. These mostly rely on the construction of a template spectrum of the integrated starlight from the host galaxy which is used to subtract the stellar continuum spectrum from the pure emission-line spectrum produced by the AGN (e.g., Filippenko \& Halpern 1984; Hao et al. 2005). The spectroscopic properties analyzed in this work come from the reduction of SDSS spectra performed by Kauffmann et al. (2003) where a very careful stellar subtraction was obtained by fitting emission-line-free regions of the spectra with model galaxy spectra obtained from the new population synthesis code of Bruzual and Charlot (2003). Such careful stellar subtraction allows the selection of large samples of AGN with apparently unremarkable lines in their original spectra.

Assuming a spectroscopically fair sample of galaxies, the selection of AGN and star-forming galaxies is then based on the relative strength of permitted and forbidden emission lines: AGN tend to display relatively strong low-ionization metal lines such as [OII], [OIII], [NII], and [SII] compared to star-forming galaxies as a consequence of their harder continuum radiation which extends into the extreme UV and X-ray portions of the electromagnetic spectrum \cite[]{Baldwin1981, Veilleux1987, Kauffmann2003, Kewley2006}. 

As an alternative to spectroscopic selection, AGN can also be selected on the basis of their X-ray and radio properties, although the current sensitivity of X-ray and radio surveys allows access only to the bright end of the AGN luminosity function in these bands.  Agreement among different methods of selection is far from perfect: in particular, the photometric, spectral, and radio classification of AGN do not agree -- large fractions of spectroscopically unremarkable galaxies show strong radio or X-ray emission \cite[]{Reviglio2006}. This might either reflect a strong difference in the physical properties of the different AGN populations, the effect of selection biases in different bands, or a combination of the two. 

In this paper we focus on the comparison of spectroscopic, X-ray, and radio selection of AGN, with the aim of disentangling true physical differences in the AGN population and their environments from differences resulting from incompleteness in selection and classification methods. The structure of the paper is as follows: in section \ref{database} we present the multiwavelength database created to collate optical, radio, X-ray, and FIR information for more than 150,000 galaxies drawn from the SDSS (DR2), while in section \ref{density} we describe the three-dimensional density estimator used to quantify the density of the environment surrounding each galaxy. In section \ref{puzzles}, we present two puzzles identified in the data when comparing radio and spectroscopic classifications of active galaxies, and then explore (\S \ref{dilution} and \S \ref{dilution_SFL}) the role of dilution in creating these puzzles. In section \ref{solution} we propose a statistical method to solve this problem and select a fair sample of active galaxies. We then analyze the effect of spectroscopic bias on the total fraction (\S 8) and spatial distribution (\S \ref{environment}) of AGN in the local universe and confirm the existence of a class of spectroscopically unremarkable AGN with X-ray and/or radio emission, the latter mostly associated with jets. We present our conclusions in section \ref{conclusion}.

\section{The multiwavelength database}                      
\label{database}

Galactic activity can be detected in several different observational bands.
In this paper we focus on the radio continuum at 1.4 GHz, the far-infrared at 60 and 100 $\mu m$, and the X-ray band. The physical mechanisms that lead star-forming galaxies and AGN to emit in each of these bands are still somewhat uncertain. For star-forming galaxies, radio emission is believed to originate in the interaction of cosmic rays, accelerated to relativistic speeds in supernova shocks \cite[]{Fermi1949}, with galactic magnetic fields, while the far-infrared emission arises from re-radiation of dust heated by high-mass, newly born stars \cite[]{Condon93}, and X-rays come from accreting binaries \cite[]{Grimm2003}. For AGN, the radio emission is supposed to come from the interaction of relativistic particles and magnetic fields in the central region, the far-infrared from heated dust in the nuclear region (typically at higher temperatures than for star-forming galaxies), and X-rays from the accretion of gas onto the supermassive black hole which powers the AGN and/or from relativistic jets \cite[]{Ferrari1998}. An indirect signature of galactic activity is the presence of strong optical emission lines arising from recombination of ionized species -- in particular, H$\alpha$, H$\beta$, [OII], [OIII], [NII] \cite[]{Osterbrock2006}.

In order to explore the properties of active galaxies we have built a new multiwavelength database considering all the galaxies from the spectroscopic sample of the SDSS \cite[]{York2000} Second Data Release (DR2) with extinction-corrected Petrosian R magnitudes in the range $14.5\le R \le 17.5$. This magnitude cut avoids the faintest objects, ensuring only higher signal-to-noise spectra are used in our study. For these galaxies we have adopted the spectral classification of Kauffmann et al. (2003), which is a refined version of the standard BPT classification scheme for emission-line galaxies (Baldwin et al. 1981). R-band magnitudes have been K-corrected using the procedure of Blanton (version v3\_2 -- \cite{Blanton2003}). The final sample includes 151,815 galaxies: 40,040 galaxies are classified as star-forming (SFG), 27,789 as weak signal-to-noise star-forming (SFL), 16,017 as composite systems (COM), and 23,702 as Active Galactic Nuclei (AGN), while 44,267 are left unclassified: this latter class includes low signal-to-noise systems with very weak or no lines \cite[]{Brinchmann2004}. We shall refer to this class as absorption-line galaxies (ABG).

Details of the classification system are given in \cite{Brinchmann2004}. We note that a minimal requirement adopted by those authors for classification as an AGN is [NII]6584/H$\alpha$ $>$0.6 and signal-to-noise (S/N) in each line $>3$. The SFL class includes all galaxies with S/N$>2$ in H$\alpha$ that were not classified as SFG, Composite, or AGN. 

Broad emission-line AGN are not included in the Kauffmann et al. (2003) database, but they account for only $\sim$ 1\% of the sample. We extracted a sample of broad-emission line AGN from the SDSS and discuss their properties in a subsequent paper. H$\alpha$ fluxes of star-forming galaxies have been corrected for aperture bias following Hopkins et al. (2003).

\begin{figure}
\begin{center}
\includegraphics[scale=0.5,angle=90]{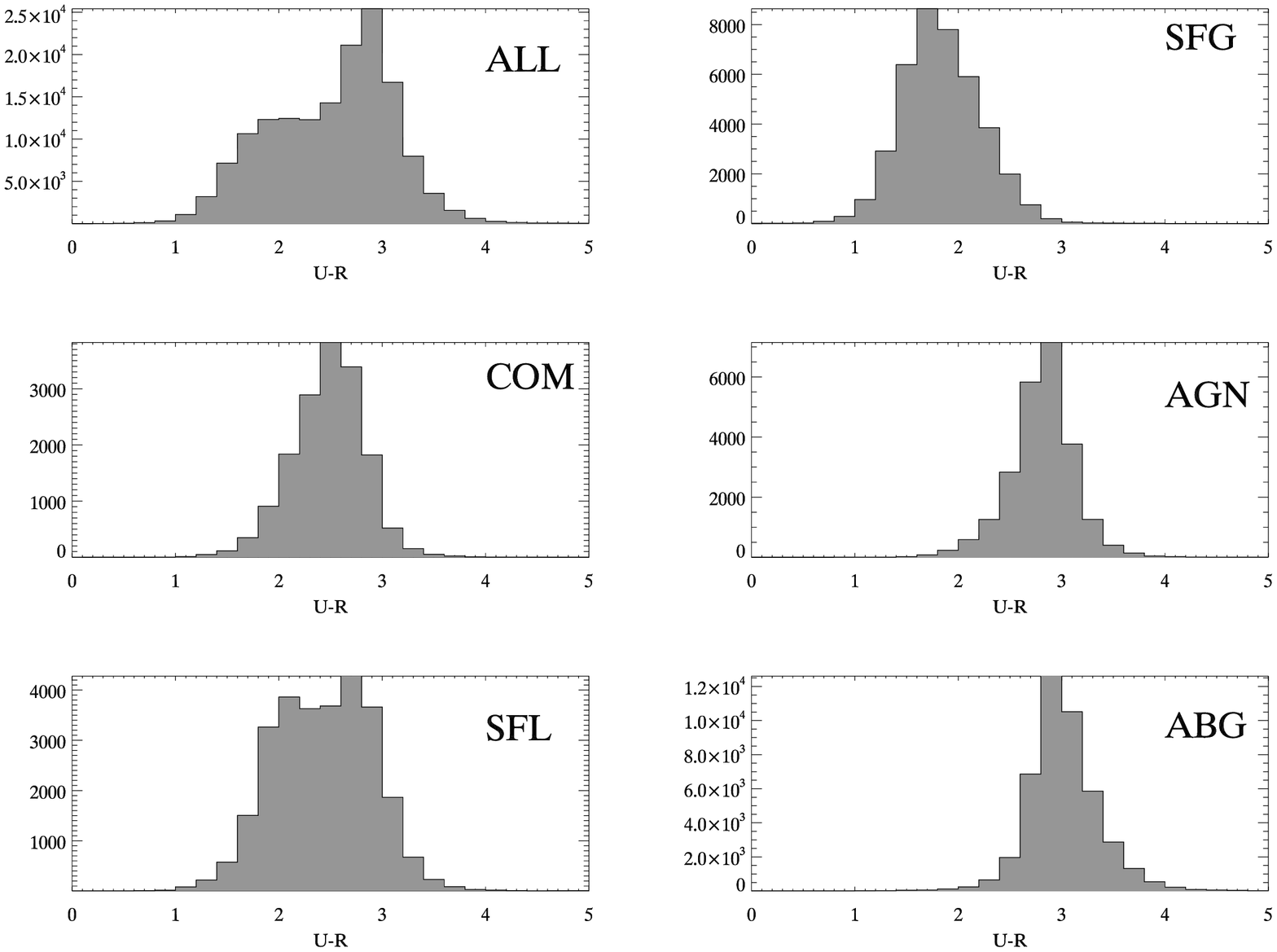}
\end{center}
\caption{\emph{Color distributions for the whole spectroscopic sample, divided by spectroscopic class:  ALL = the entire $\sim 150,000$ galaxy sample, SFG = star-forming galaxies, COM = composite objects, AGN = emission-line AGN, SFL = low signal-to-noise star-forming galaxies, and ABG = absorption line galaxies.}}
\label{histo_color}
\end{figure}

We obtained a crude division of these galaxies into early-type (bulge-dominated) and late-type (disk-dominated) by following the prescription of Strateva et al. (2001). Their study showed that the best way to separate early-type galaxies from late-types is to consider $u-r$ color: galaxies with $u-r>2.22$ are early-type with 98\% completeness and 83\% reliability, while galaxies with $u-r<2.22$ are late-type with 72\% completeness and 96\% reliability. In figure \ref{histo_color} we present the distribution in color for our spectroscopic sample. We have used the cut in color at 2.22 and find 86,311 early-type galaxies and 65,504 late-type galaxies. We further checked the consistency of this coarse classification by comparing it with the information available from the surface-brightness profiles of these galaxies. Since the surface brightness profiles of early-type galaxies are known to be well-described by a deVaucouleurs profile while for the disks of late-type galaxies the profile is better approximated by an exponential, we compare the likelihood of the two profiles for each group of galaxies in our sample. We find that 85.1\% of our early-type galaxies have the likelihood of a fit to the de-Vaucouleurs profile $L_{dV}$ higher than the likelihood of the exponential disk profile $L_{exp}$, while for the late-type galaxies, 73.2\% have $L_{exp}>L_{dV}$. Blue ellipticals which, in this classification scheme may be misclassified as late-type systems, account only for a small percentage of the whole galaxy population ($\sim$ 3\% according to \cite{Driver2007}). 

We cross-correlated our sample with the NVSS \cite[]{Condon1998} and FIRST \cite[]{Becker95} radio catalogs, the IRAS Point Source (IPAC 1986) and Faint Source catalogs \cite[]{Moshir90}, and the ROSAT Bright Source \cite[]{Voges99}, Faint Source \cite[]{Voges00}, and WGACAT catalogs \cite[]{White00}. In order to define the optimal matching radii, we built for each catalog four false catalogs by offsetting all positions $1^{\prime}$ in each of the four cardinal directions and calculating the number of false matches expected for different search radii. The final search radius $R_s$ was chosen to strike a balance between having the largest number of real matches and a modest number of spurious ones; the selected search radii give $\sim 90\%$ completeness and $\sim 90\%$ reliablility for all surveys.  When more than one source is associated with an optical galaxy, we assign the closest match to the source.

For FIRST, the search radius adopted is 6'', while for NVSS it is 20''; we find 6075 matches with FIRST and 5480 matches with NVSS. Matches with NVSS without a match in FIRST have been checked for the presence of radio sources detected by FIRST within the annulus between 6'' and 20''. This allowed the removal of false matches arising from the larger NVSS search radius.

The total number of unique radio matches is 8058; roughly one-fourth of the total radio sources are detected only in NVSS, but not in FIRST and  one-third of the total radio sources are detected only in FIRST. For the first group we find that 41\% fall in areas of the sky covered by SDSS-DR2 and NVSS but not by the FIRST Survey, while the remaining sources of the first group  are mostly low-surface-brightness extended sources that are resolved out in FIRST because of its lack of short interferometric spacings. In agreement with this interpretation, more than half of these sources are classified as star-forming or composite systems, a fact that suggests that radio emission is extended and therefore is more readily resolved out by the FIRST survey. The vast majority of the sources in the second group are below the detectability threshold of NVSS (2.5 mJy) but above the detectability threshold of FIRST (1 mJy). 

In order to obtain an independent classification of the activity producing the radio emission in our radio-bright sub-sample, we employed a method similar to that described in \cite{Reviglio2003, Reviglio2006, Reviglio2008}, which takes full advantage of the optical and radio morphological information available for our sources. Given the high resolution of the FIRST Survey, we are able to divide our radio sources into two categories: unresolved (angular size $\leq $2.5'') and extended (angular size $>$2.5''). Broadly speaking, early-type galaxies with extended radio emission have been classified as jets (JET), while early-type galaxies with unresolved (point-like) emission have been classified as galaxies with an active galactic nucleus (ACT). Late-type galaxies with extended radio emission are classified as star-forming galaxies (SFG).

We added several other criteria to refine this simple classification scheme. Since our classification into early- and late-type galaxies is given by a sharp cut in color, we expect a fraction of redder star-forming systems to be misclassified as early type systems -- in particular, the edge-on spirals for which intrinsic extinction may affect their observed color. For the scheme presented, these star-forming systems, when detected in the radio, would be misclassified as JET sources. To avoid this problem, we compare the alignment of the radio emission with the galaxys' optical axes, since in star-forming galaxies the orientations tend to be aligned, while in galaxies powering radio jets they do not. We analyzed the distribution of the relative position angles for these two classes of objects and, upon inspection of a sample of images, reclassified all systems with color $u-r$ bluer than 2.9 and position angle offsets smaller than 25$^\circ$ as star-forming, since most appear to be reddened spirals with ongoing star formation. The remaining (redder) color cut is necessary to avoid misclassification of ellipticals.

We left unclassified the late-type galaxies with point-like radio emission since they might be either nuclear starbursts or Seyfert galaxies. We kept track of the point-like sources by labeling them PNT. When radio sizes were not available (NVSS sources without a FIRST counterpart), we left them unclassified (UNC); these objects were ultimately classified according to their optical spectra, as discussed below.

In order to account for the large resolved jets (mostly FR I and FR II sources) which are generally cataloged as several discrete sources in the FIRST Survey,
we took advantage of the fact that most radio jet sources in the \emph{local} universe ($z \lesssim 0.2$) also have detectable nuclear radio emission centered on the optical galaxy (as we found in the smaller sample of $\sim$400 radio sources examined in our previous study \cite[]{Reviglio2006}).

This fact allows us to select a short list of potential hosts of radio jets by requiring that they have an associated nuclear radio source; as noted above, our matching radius is 6$^{\prime\prime}$. The existence of multiple sources within 2$^\prime$ is then considered as an indication of the potential existence of a radio jet. A short list of 910 fields with potential large jet-powered sources is thus gleaned from an initial list of more than eight thousand matches between the two catalogs. We classified by eye the sources in these 910 fields and found 148 FR I sources (bright central emission, faint lobes), 174 FR II sources (weaker central emission and bright lobes), and 146 JET sources (asymmetric emission or no clear lobes); the remaining $\sim 50\%$ of the candidates appear to be unrelated background sources and are not considered further here.

This procedure however misses potential FR2 sources without central emission. In order to recover these sources we took advantage of the fact that the NVSS beam size is much larger than the beam size of the FIRST survey and therefore FR2s are not split into multiple sources. We  visually inspected all fields with NVSS emission and two or more FIRST sources within 20'' and recovered 96 FR2 with undetected central emission in FIRST.

After adding all these sources with extended emission to those AGN selected by our automated algorithm (early-type galaxies with extended radio emission not parallel to the optical axis), we obtain the following final classification for our radio sources: 1433 SFG sources,  1784 ACT sources, 2114 JET sources, 148 FR1 sources, 270 FR2 sources, 422 PNT sources, 1887 UNC sources. As a further sanity check, we compered the luminosity function obtained for this radio sample with the one obtained by \cite{Sadler2002} and found very good agreement.

In order to explore the X-ray properties of our sources we cross-correlated our sample with the ROSAT X-ray catalogs cited above. We selected only point-like sources, since extended X-ray emission associated with galaxies might arise from the hot intracluster medium surrounding them and would not be directly relevant to galaxy activity. We obtained 382 matches using a search radius of 30''.

All X-ray sources with an X-ray luminosity greater than $10^{42}$ erg s$^{-1}$ have been classified as AGN, as it has been shown that other sources of X-ray emission in galaxies (e.g., X-ray binaries, the hot interstellar medium) cannot account for such high X-ray luminosities \cite[]{Grimm2003}. This way we were able to independently classify 360 X-ray emitting AGN. Interestingly, the vast majority of our x-ray detections are radio-emitting AGN and the x-ray classification is in excellent agreement with the radio classification; only 6\% of these x-ray AGN with radio emission  are not classified as radio AGN. There are, of course,  AGN with X-ray luminosities below $10^{42}$ erg s$^{-1}$, and some of the 22 low-L$_X$ may also belong in this category, but in order to minimize contamination of the AGN sample, we do not include them here, as they represent a small fraction of the total. We do find that five of these X-ray sources with luminosities below $10^{42}$ erg s$^{-1}$ are classified as AGN because of their radio or spectroscopic features. 

Finally, we cross-correlated our optical sample with the IRAS Point Source (IPAC 1986) and Faint Source catalogs \cite[]{Moshir90}, using a search radius of 40'' determined in the manner described for the radio surveys; we find 3152 matches.

\section{The density estimator}
\label{density}

To associate a volume density with each galaxy, we have implemented the 
procedure outlined in Carter et al. (2001) for magnitude-limited samples. The idea of this method is to calculate the relative densities of different environments by considering the volume enclosing the ten nearest \emph{observed} neighbors of a given galaxy j and then dividing by the number of galaxies estimated to be present in the volume by the volume itself $V_j$. In order to find the number of galaxies $N_j$ actually present in $V_j$, it is necessary to correct the number observed in the volume (ten, by definition) to account for galaxies which fall below the magnitude limit of the sample. Assuming that the Schechter optical luminosity function is indeed universal (changes in the functional form in different environments are not too large -- e.g., Christlein 2000), the correction can be obtained by considering that, in order to have one galaxy with absolute magnitude M in a unit volume, there must be a certain number of galaxies, as defined by the luminosity function, which are less bright in the same volume. 

The estimate of the environment density for the $j^{th}$ galaxy is, then:
\begin{equation}
\rho_j=\frac{N_j}{V_j}
\end{equation}

where $N_j=\displaystyle\sum_{i=1}^{j}\frac{\int_{-\infty}^{+\infty} \phi(M)d(M)}{\int_{-\infty}^{M_i} \phi(M)d(M)}$ with $M_i$ the faintest absolute magnitude detectable in the survey at the $j^{th}$ position and $\phi(M)$ is the Schechter luminosity function.  Since the absolute magnitude of galaxies effectively ranges not from plus to minus infinity, but between a maximum and minimum value, we have adopted the maximum (R=-13)and minimum (R=-25)absolute reliable magnitudes in the survey to calculate the integrals.  For galaxies close to the survey borders, we have corrected the estimate of the density by evaluating the number density of galaxies inside the fraction of the volume given by the intersection of their spheres with the survey borders. We have used an R-band luminosity function with Schechter parameters $\phi_*=0.005 \rm{Mpc^{-3}}$, $\alpha=-1.17$, $M_*=-20.06$ as determined by Blanton et al. 2001 in order to perform the evaluation of the number density of galaxies surrounding a given galaxy. Luminosity distances $D_L$ have been calculated using standard cosmological parameters ($\Omega_0=1$ , $\Omega_M=0.3$, $\Omega_\Lambda=0.7$.and $H_0=70 \rm{km \ s^{-1}Mpc^{-1}}$ -- Spergel et al., 2003), and angular distances $D_A$ have been calculated using the standard general relativistic scaling between angular distances and luminosity distances, $D_A=D_L/(1+z)^2$. From Poisson statistics, the densities are determined with a $\sim$ 30\% relative error, so the uncertainty in the logarithm is $log(\rho)\sim 0.1$; we thus always bin our data with $\Delta Log(\rho) \ge 0.5$ to make sure that smearing and the small degree of incompleteness in the survey do not bias our results.

The method employed samples densities on a few megaparsec scale and it is therefore a measure of the large-scale environment (voids, filaments, clusters) in which galaxies reside. All density estimators suffer of systematic uncertainties.
One of the major strengths of this method is that it allows us to associate an actual three-dimensional density with all galaxies belonging to the magnitude-limited sample under study, effectively allowing us to study the environment of a very large number of objects, while minimizing the systematics of other methods, such as counting galaxies in a fixed volume, that might lead to an overestimate or underestimate of over-dense and under-dense environments. A more complete discussion of the method can be found in Carter et al. (2001).

The SDSS spectroscopic sample is 85\%-90\% complete to the spectroscopic limit but has a 7\% incompleteness given by fiber collisions which are preferentially associated with dense cluster cores; this leads to an under-sampling of the densest regions. This could bias the density estimate of the highest density bin, although this is not a severe problem in our analysis since we utilize fairly large bins.  For example, for  $\rho$=10 galaxies/Mpc$^3$, an incompleteness of 10\% would lead to an estimate of $\rho$=9 galaxies/Mpc$^3$ and the difference in the logarithms of these two densities is $\sim$ 0.05, smaller than the Poissonian error for this method. Moreover, Miller et al. (2005) have suggested that this under-sampling is modest in rich clusters; they find that $>$70\% of the rich cores of Abell clusters in the SDSS are at least 70\% complete. We also note that the distribution in density we find is similar to that obtained in our earlier paper (Reviglio \& Helfand 2006) for the 15R survey \cite[]{Carter01} which has different systematics than the SDSS and is unaffected by fiber- collision incompleteness. We therefore conclude that our current sample includes a fair sampling of densities. We note that, unlike many studies, we are using actual three-dimensional densities, not projected densities. A comparison with studies using other density estimators must be done with care because of the different systematics affecting each method. 

\begin{figure}
\begin{center}
\includegraphics[scale=0.5,angle=90]{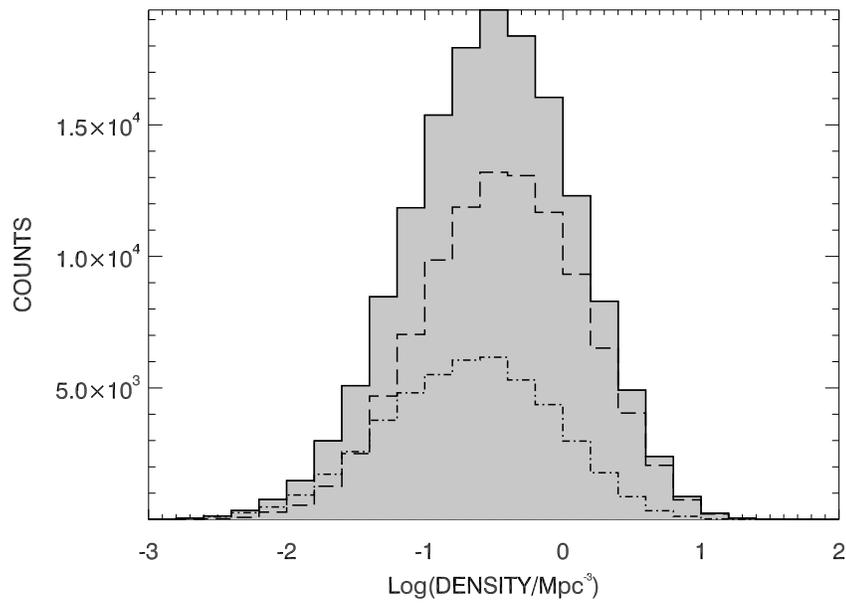}
\end{center}
\caption{\emph{The local density distribution for galaxies in our sample with $z < 0.2$ . The dashed line shows the distribution for the early-type sample, while the dot-dashed line shows the distribution for the late-type sample. As expected, early-type  galaxies dominate the fractional abundance in denser environments, while late-type galaxies predominate in lower density regions.  }}
\label{histo_density}
\end{figure}

A histogram of the distribution of the environmental densities for our sample of galaxies is given in figure \ref{histo_density}. High- and low-density environments account for just few percent of the distribution.  Early-type (or equivalently, red-sequence) galaxies and late-type (blue-sequence) galaxies are found in all environments, with the relative abundance of early to late types increasing steeply with increasing density, as expected. 

\section{Puzzles with the spectral classification}
\label{puzzles}
Reviglio and Helfand (2006) have shown that the spectroscopic, morphological, and photometric classifications of a radio-emitting sub-sample of spectroscopically selected AGN do not agree.  In this section we re-examine the problem of the mismatch between the optical and radio classification of galaxies. The wealth of information available from the SDSS with its much larger galaxy sample allows us to clarify the nature of the mismatch and to explore its implications for the study of samples of active galaxies selected spectroscopically.

The spectral classifications for the galaxies that show radio emission are as follows: 1893 SFG sources, 377 SFL sources, 1438 COM sources, 1866 AGN sources, 2484 ABG sources. For X-ray selected AGN (x-ray luminosity $> 10^{42}$ erg s$^{-1}$) we find: 27 SFG, 29 SFL, 49 COM, 86 AGN, and 169 ABG.

From these data two main puzzles arise:

\begin{itemize}

\item \emph{44\% of galaxies classified as active galactic nuclei from their radio emission are classified as passive systems based on their spectra.}

These galaxies preferentially exhibit jet-like radio structures: 74\% are classified as jets and only 24 \% as point-like radio sources in early-type galaxies; the rest are mostly sources with no available radio morphology. Interestingly, 70\% of our FR I sources and 77\% of our FR II belong to this class of optically passive objects. Similarly 47\% of X-ray AGN are classified as passive galaxies. These results are consistent with our results from the smaller 15R Redshift Survey \cite[]{Reviglio2003, Reviglio2006} and with studies by other authors \cite[]{Martini2002, Best2004}, suggesting that the problem might reside in the use of emission lines as the exclusive diagnostic to classify galaxy activity, rather than with the systematics of each survey.

\item \emph{More than 50\% of SFL galaxies with radio emission are classified as AGN according to their radio morphologies and optical colors.}

Only 2\% of the SFL galaxies with colors redder than $u-r = 2.2$ have been classified as SFG galaxies according to their radio properties, while more than 60\% of these galaxies have been classified as AGN in the radio: 40\% show jet-like emission (22\% of these as FR1 or FR2 objects) and 20\% show point-like AGN emission. The rest are unclassified systems.  When the X-ray sample is considered, 8\% of the X-ray AGN are classified as SFL of which 86\% have colors redder than 2.2. 

\end{itemize}

In the next section we will explore these problems, evaluating whether these discrepancies between classification schemes arise from actual physical processes or are a result of survey systematics. 

\section{The problem of dilution for active galactic nuclei}
\label{dilution}

A fair classification of galaxies in the local universe must be invariant over the volume sampled, provided that a) the volume is larger than the homogeneity scale  of the universe, b) the redshift range is small enough to avoid strong evolutionary effects in the composition of the sample, and c) the optical sample is complete. Despite the flux-limited nature of the SDSS sample, it is possible to select volume-limited sub-samples of galaxies in order to compare galaxy classifications. Volume-limited selection ensures that a complete sub-sample of galaxies in a given luminosity range is selected, since it limits the galaxies considered to the sub-sample of objects with absolute magnitudes visible at all redshifts within the chosen volume. 

One way to evaluate whether or not a certain classification scheme for galaxies is unbiased is to compare the relative numbers of galaxies in each spectroscopic class in different redshift bins using a volume-limited sample selected to have a fairly narrow luminosity range, and to check that the fraction of each galaxy class remains constant with distance. In particular, for a survey with conic geometry such as the SDSS, the volume sampled increases with increasing redshift, but the fraction of objects assigned to each galaxy class should be invariant. The stronger the deviation from constant fractions, the higher the likelihood that misclassification is involved.

\begin{figure}
\begin{center}
\includegraphics[scale=0.5,angle=90]{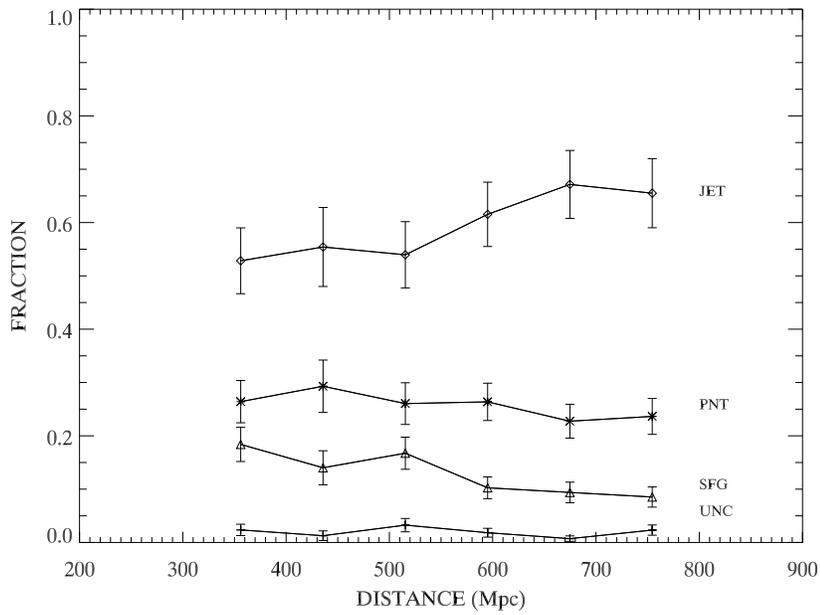}
\end{center}
\caption{\emph{The fractional abundance of different classes of radio-emitting objects as a function of distance for a volume-limited sample of galaxies defined by $-23<R<-22$. The radio classification is reasonably fair, in that the fraction of each class of galaxies is roughly constant with distance.}}
\label{dilution_counts_radio}
\end{figure}

In this section we focus on a volume-limited sample defined by $-23<R<-22$ and the associated luminosity distance cut $316~{\rm Mpc} < D_L < 794~{\rm Mpc}$, which comprises 18,048 galaxies. In figure \ref{dilution_counts_radio} we show this type of analysis for the radio sub-sample; the fractions of each of the four main classes of radio sources are plotted versus distance. The trends for the different classes are quite flat, with a small decline in the JET fraction and a small increase in the SFG fraction at lower redshifts. The trend for point-like AGN is remarkably flat. This shows that, even if a small fraction of the putatively JET-like structures have been misclassified as SFG, the radio classification is substantially correct; any misclassification would affect $\lesssim 10\% $ of the sample and would not bias our conclusions significantly. We note that, from the magnitudes of the trends, we can exclude the possibility that a significant fraction of SFG galaxies were misclassified as JET sources and not vice-versa, since if we were to add the excess of JETs at higher redshift to the SFG sample, we would obtain an increasing SFG fraction with distance. Most likely, the source of the minor misclassification is that a small fraction of early-type galaxies with radio emission centered on the optical core which we reclassified as SFG are, in fact, jets aligned with the optical major axis of the source. Since there is no way to tell the difference between these two cases by looking at the images and the degree of the misclassification is minor, we have not modified our classification scheme.

What happens, in contrast, if we consider the galaxies in the SDSS sample classified according to their optical spectra? 

\begin{figure}
\begin{center}
\includegraphics[scale=0.7,angle=90]{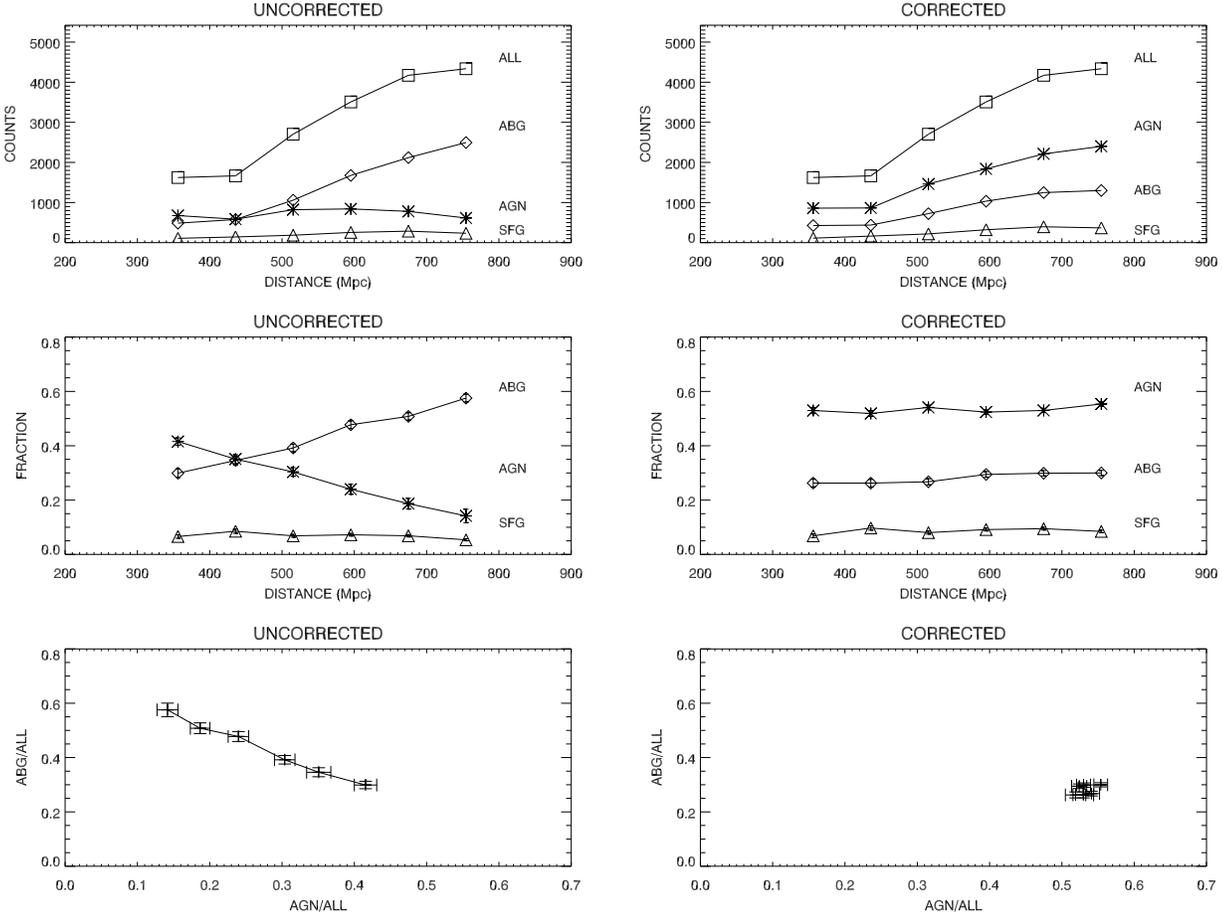}
\end{center}
\caption{\emph{On the left: the distribution of different spectroscopic classes for our sample. The top panel shows galaxy counts while the second panel shows the fractional abundance and the bottom panel shows the fraction of galaxies classified as AGN, all as a function of distance for the volume-limited sample of galaxies defined by $-23<R<-22$. The spectral classification is clearly biased. A substantial fraction of emission-line AGN must be misclassified as passive (ABG) galaxies. On the right: the trends after our correction is applied.}}
\label{dilution_counts}
\end{figure}

The results for the volume-limited samples are shown in the left column of figure \ref{dilution_counts}. In this case, the central panel shows strong trends in the fractional abundances of emission-line and passive systems with distance. Emission-line AGN are four times more abundant in the lowest redshift bin than in the highest one. It is clear that absorption-line systems are over-represented at higher redshifts, while the emission-line AGN are under-represented. The classification is clearly biased and the trends suggest that a fairly large fraction of emission-line systems have been misclassified as passive galaxies.  This might well explain the large fraction of radio AGN associated with passive galaxies noted above.

Star-forming galaxies, by contrast, have a fairly flat distribution, suggesting that their classification is reasonably fair.  A minor decrease in the fractional abundance of star-forming systems is found (see figure \ref{dilution_counts_SFL} where this trend is more apparent), but it is much milder than the trend seen for the AGN. Some of the SFG galaxies at higher redshift must, however, also have been misclassified. 

The simplest explanation for the misclassification of distant emission-line systems is the dilution of their lines by light from the host galaxy which represents an increasing contribution to the light entering the fiber as its apparent size becomes smaller. As noted earlier, dilution is known to affect the classification of Seyfert galaxies: \cite{Moran02} have shown that when the size of a Seyfert galaxy approaches the size of the slit used to obtain the spectrum of its nuclear region, the emission lines produced by the AGN are diluted by the light of the host galaxy falling inside the slit and are not detected. Hao et al. (2005) note the importance of this effect in their derivation of the luminosity function for a sample of SDSS AGN.

\begin{figure}
\begin{center}
\includegraphics[scale=0.5,angle=90]{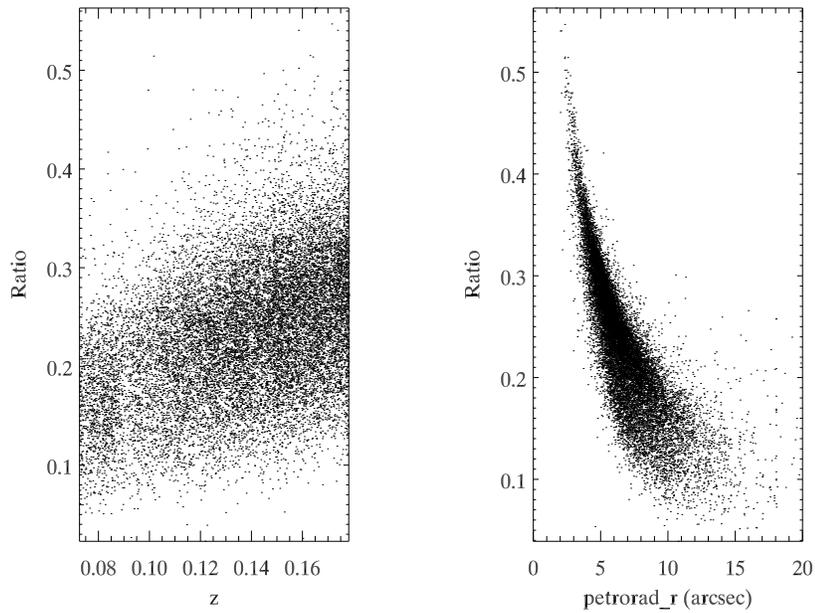}
\end{center}
\caption{\emph{On the left: the increase in the ratio of r-band light entering the fiber with increasing redshift for all galaxies in the $-23<R<-22$ volume-limited sample . On the right: the increase in the fraction of r-band light entering the fiber for galaxies with smaller and smaller observed sizes for the same sample. Sizes are evaluated using R-band Petrosian radii.}}
\label{dilution_mag_ratios}
\end{figure}

For all galaxies in the volume-limited sample considered here, the plots in figure \ref{dilution_mag_ratios} quantify the increase in the fraction of galaxy light that falls inside the SDSS fibers with increasing  redshift and decreasing apparent size by considering the ratio of the R-band flux falling inside the fiber to the flux of the whole galaxy. In general the increases are more significant for early-type systems since a de-Vaucouleurs light profile is more concentrated than the profile of an exponential disk. In the volume-limited sample considered here, early-type systems form the large majority, given the high luminosity threshold we have selected to explore this problem.

If dilution is responsible for the misclassification of the spectroscopic sample, we should find that the relative abundance of absorption-line systems to emission-line systems must increase not only with increasing redshift, but also with decreasing observed host size, since galaxies with smaller observed size have a larger fraction of their light entering the fiber (see below).

\begin{figure}
\begin{center}
\includegraphics[scale=0.5,angle=90]{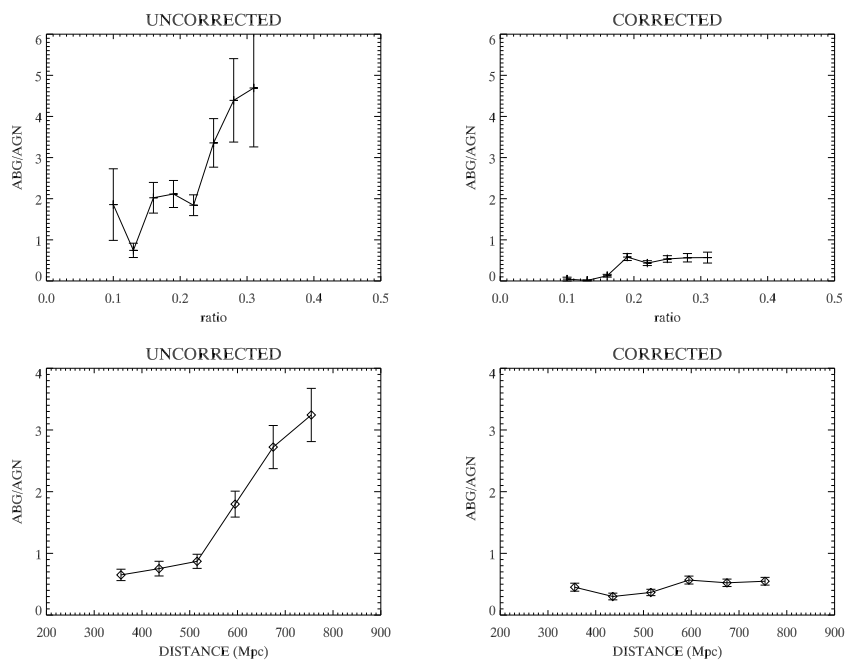}
\end{center}
\caption{\emph{On the left: the ratio of radio AGN without detected emission lines (ABG) to radio AGN with detected emission lines as a function of the ratio of the galaxy light falling into the slit(top panel) and as a function of distance (bottom panel) for our volume-limited sample of radio-emitting galaxies with $-23<R<-22$. The strong trends show that the spectral classification is biased, most likely as the result of dilution by host galaxy light. On the right: the trends after our correction is applied.}}
\label{dilution_radio_AGN}
\end{figure}

One potential problem with our approach is that some fraction of the spectroscopically passive galaxies are truly passive and their inclusion will bias the analysis. To avoid this, we run our test on a sample of radio AGN, appropriately chosen in order to ensure that the same type of host is picked at all distances and sizes. We adopt several tight cuts in absolute magnitude and color, and plot the fraction of ABG galaxies to radio AGN against both size (as traced by the R-magnitude Petrosian radius) and redshift. In figure \ref{dilution_radio_AGN} we show the trends for a volume-limited sample of 1185 radio AGN with host galaxy color $3<u-r<4$ and absolute R magnitude $-23<R<-22$. For this choice of magnitude cut, the volume selected falls near the median survey redshift. A strong increase with increasing redshift and a strong increase with increasing fraction of the light falling into the fiber are evident. The plot shows a net increase of a factor $\simeq 4$ in the abundance of absorption-line systems when the distance doubles. This strongly suggests a major selection effect in the spectroscopic selection of AGN: more and more emission-line systems are misclassified as absorption-line galaxies as the distance increases. Similar results are found if the range in color is further narrowed to just include galaxies with $3.0<u-r<3.2$, demonstrating that the trend is not the result of a varying composition in the host population with redshift in our volume-limited sample (such as an increase in the elliptical to spiral fraction). The cuts in luminosity and color ensure that similar types of hosts are considered in each redshift bin. The median mass of the hosts does not vary significantly with varying distance, nor does the median bulge-to-disk ratio.  We also note that the environment does not play a role in creating this trend, since we find that the average local density in each bin is roughly constant.

One might wonder about the role of dust extinction in suppressing the emission lines, since narrow-line AGN are known to be heavily extincted systems: a thick dusty torus surrounding the broad-line region of an AGN can obliterate the Balmer emission from the broad line region. However, we note that the spectroscopically unremarkable sources do not show the [OIII] line either, a line which originates in more distant regions of an AGN and is not expected to be as heavily extincted as the Balmer lines. We also note that dust-extinction is not a redshift-dependent quantity (at least for the small redshift range we are examining) and thus is not helpful in explaining the increase in misclassified systems with redshift.

\begin{figure}[h]
\begin{center}$
\begin{array}{cc}
\includegraphics[width=2.5in]{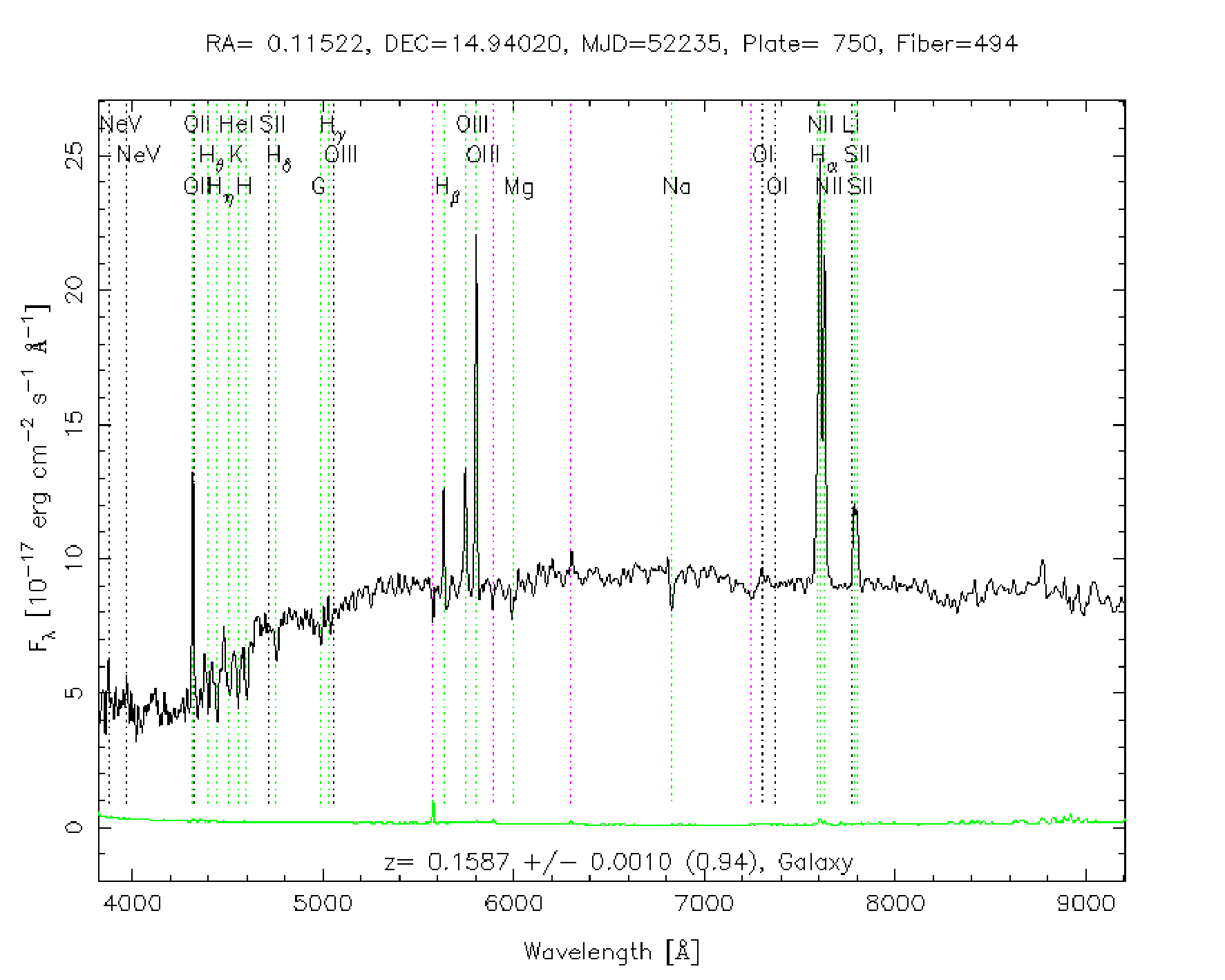} &
\includegraphics[width=2.5in]{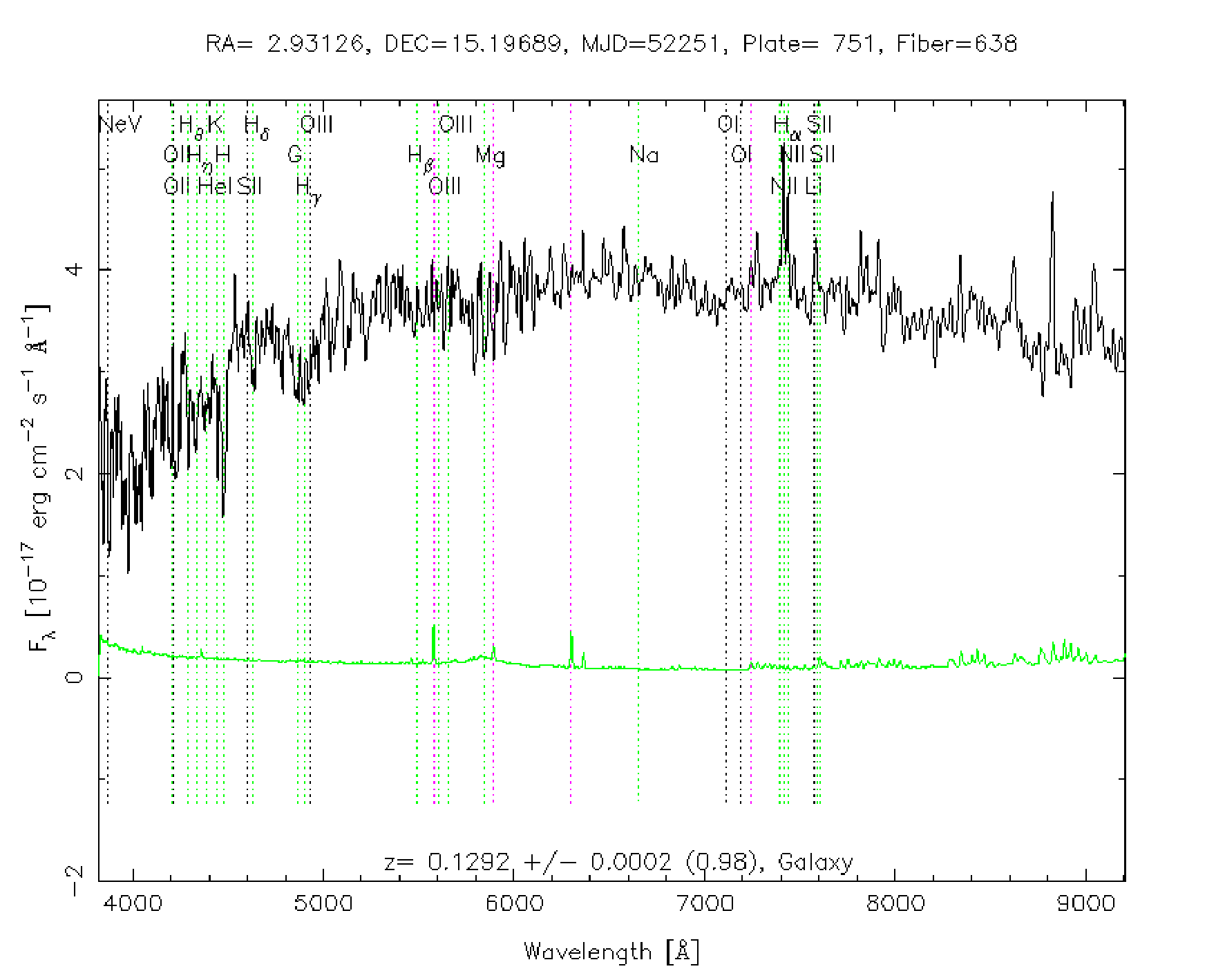} \\ 
\includegraphics[width=2.5in]{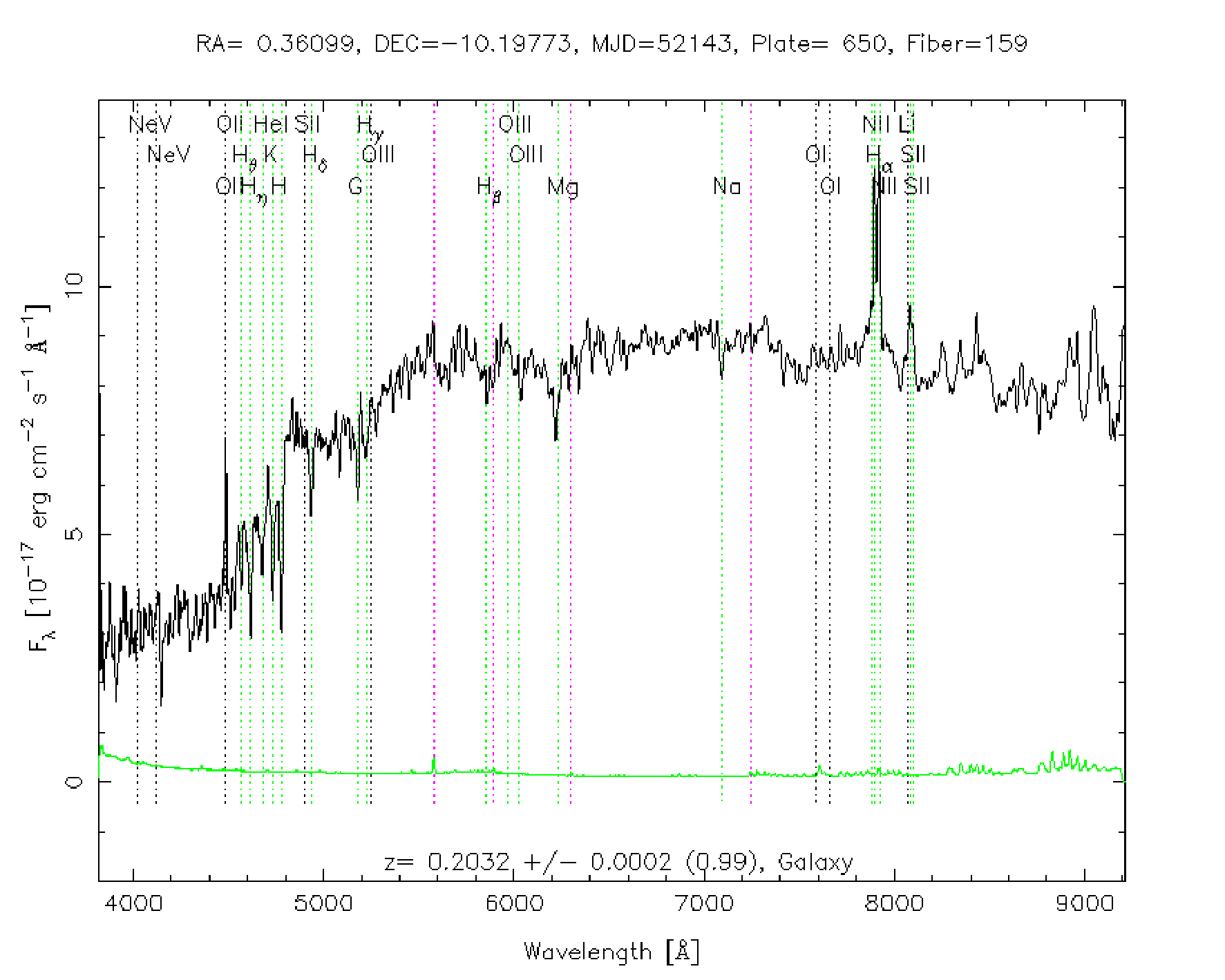} &
\includegraphics[width=2.5in]{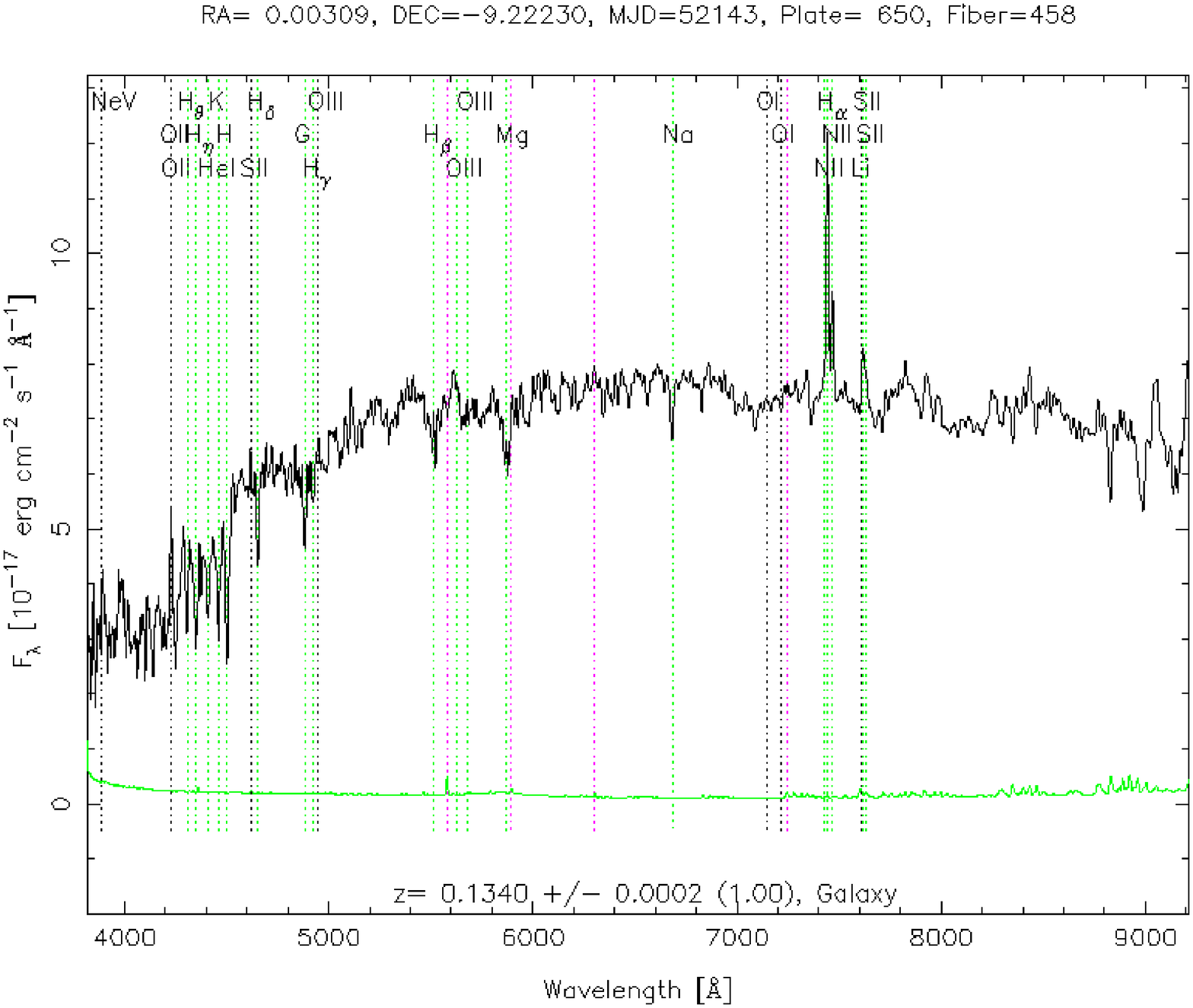}
\end{array}$
\end{center}
\caption{\emph{Clockwise from upper left corner: an example of a galaxy with a standard AGN signature, an example of a galaxy with low signal-to-noise H$\alpha$ or [NII] lines left unclassified (ABG), an example of a galaxy lacking H$\beta$ and [OIII] and classified as SFL, and  an example of a galaxy lacking H$\beta$ and [OIII] and classified as an AGN. }}
\label{spectra_sample}
\end{figure}

Dilution does not affect all AGN types in the same way. The population of emission-line AGN discussed above includes both galaxies where all standard BPT lines (H$\alpha$, H$\beta$, [NII] and [OIII]) were detected and galaxies where only H$\alpha$ and [NII] were reliably detected. Examples of the spectra of each class are presented in figure \ref{spectra_sample}. If we select all galaxies where all lines were reliably detected by requiring their equivalent widths to be $>1.0$, we find a subsample of 5212  galaxies which accounts for only $\sim$ 22\% of the total number of AGN at z$<0.2$; the large majority of AGN in the local universe do not show strong H$\beta$ or [OIII].

\begin{figure}
\begin{center}
\includegraphics[scale=0.6,angle=90]{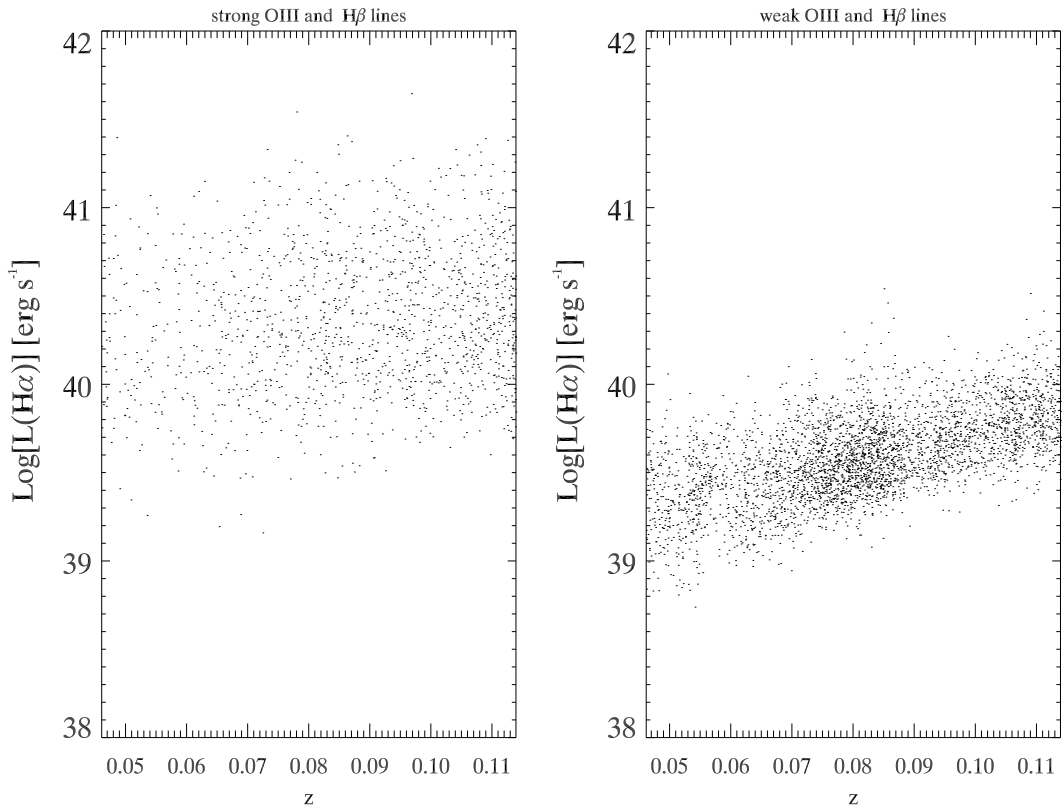}
\end{center}
\caption{\emph{On the left: for galaxies with strong H$\beta$ and [OIII] lines we do not find a significant lack of sources at higher redshifts. On the right: for systems with weak or missing H$\beta$ and [OIII] lines, a severe lack of systems at higher redshifts is present as a consequence of the increasing dilution of nuclear lines with distance.}}
\label{OIII_hb_z_AGN}
\end{figure}

In figure \ref{OIII_hb_z_AGN} we show that, for our volume-limited sample of galaxies with -23$<$R$<$-22, systems in which  all lines are detected do not exhibit significant incompleteness with increasing redshift: we do not find a rising lower cut in the scatter plot for the H$\alpha$ luminosity with increasing redshift.

\begin{figure}
\begin{center}
\includegraphics[scale=0.6,angle=90]{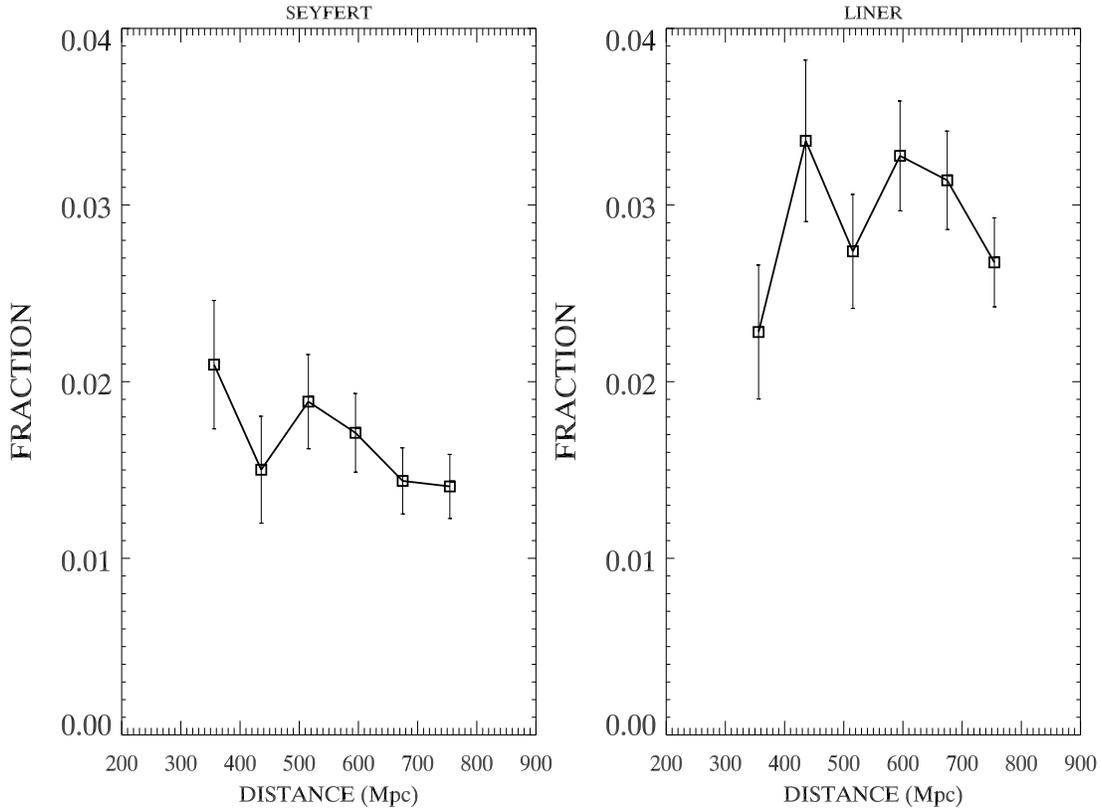}
\end{center}
\caption{\emph{The fractional abundance of Seyferts and LINERs with all lines detected  in a volume-limited sample with $-23<R<-22$  does not decrease substantially with distance. The misclassification mostly affects systems where [OIII] and H$\beta$ are not detected.}}
\label{dilution_sey_lin}
\end{figure}

If we divide the members of this class into Seyfert II galaxies and LINERs adopting the standard division among these two classes ( $\frac{EW([NII])}{EW(H\alpha)} > 0.6$ and $\frac{EW([OIII])}{EW(H\beta)} > 3$ \cite[]{Ho2004}),  we find that Seyfert galaxies account for $\sim$ 36\% and LINERS for $\sim$ 64\%, in agreement with the results of Ho et al. (2003) from the Palomar Survey \cite[]{Filippenko1985, Ho2003}, a high-sensitivity, small-slit (and much smaller) spectroscopic survey of nearby active galaxies, where the effects of dilution and pollution are minimized. We also find that the fractional abundance of the two populations does not vary significantly with increasing distance in a volume-limited sample, as shown in figure \ref{dilution_sey_lin}.

On the contrary, if we select in the $-23<$R$<-22$ volume-limited sample the population with unremarkable H$\beta$ and [OIII] (EW in these lines $<$ 1) which are mostly classified using the [NII]/ H$\alpha$ ratio, severe  incompleteness appears, as shown in the right panel of figure \ref{OIII_hb_z_AGN}. This LINER-like class of objects with particularly low [OIII] and H$\beta$ is the sample that becomes misclassified with increasing redshift because of dilution. Since this class represent the vast majority (78\% in our sample) of active nuclei in the local universe, an estimate of the magnitude of the problem is warranted.

\section{The problem of dilution for composite systems}
\label{dilution_SFL}

\begin{figure}
\begin{center}
\includegraphics[scale=0.5,angle=90]{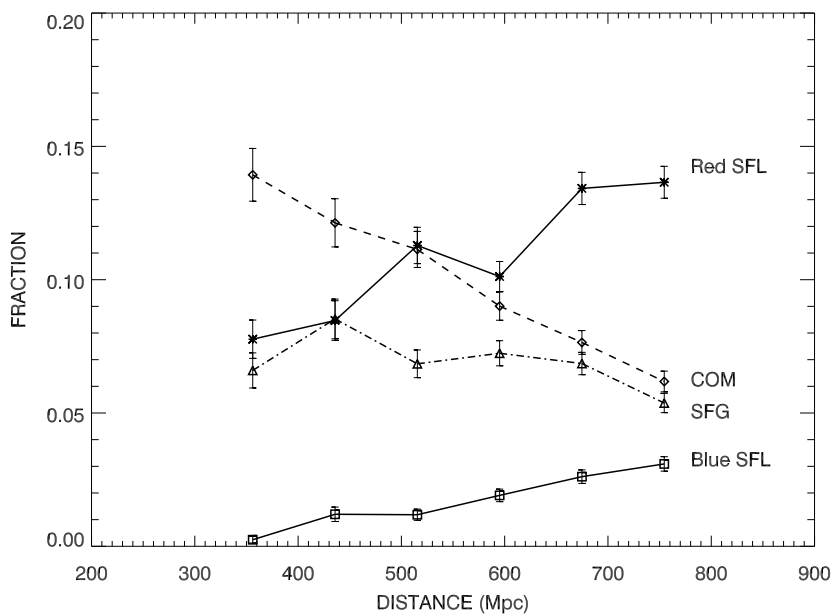}
\end{center}
\caption{\emph{The fractional abundance as a function of distance for spectroscopically classified SFG systems (triangles), red SFL systems (stars), blue SFL systems (squares) and composite systems (diamonds). The dividing color for the SFL galaxies has been chosen as $u-r=2.2$. The plots show evidence for misclassification of the SFL systems as well. Adding blue SFL systems to the SFG sample would completely flatten the trend for star-forming galaxies. Red SFL systems would compensate for the lack of composite systems at higher redshift.}}
\label{dilution_counts_SFL}
\end{figure}

We next consider SFL galaxies, the class of low signal-to-noise, putatively SFG systems, which gathers all systems with S/N in H$\alpha>2$ which are not classified as AGN, composite, or SFG. In figure \ref{dilution_counts_SFL} we find an increase with redshift that is greater than the growth required to flatten the small decline seen among SFG systems. Moreover, as reported in the previous section, most radio-emitting SFL objects appear to be associated with radio AGN. Some misclassification must be present in  this class of objects as well.

\begin{figure}
\begin{center}
\includegraphics[scale=0.5,angle=90]{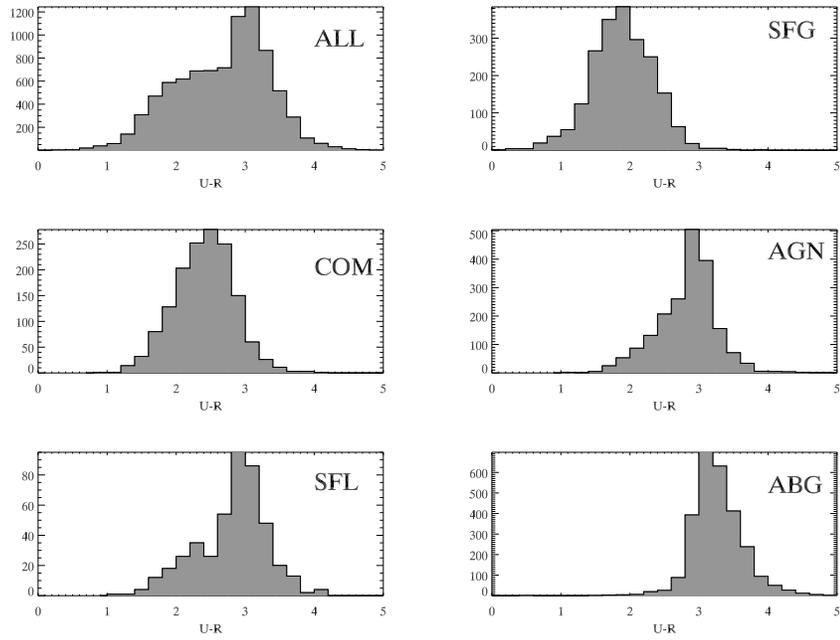}
\end{center}
\caption{\emph{Color distributions of the radio-detected sample; labels as in figure 1.}}
\label{histo_color_rad}
\end{figure}

If we consider the distribution in color for the whole sample of SFL galaxies, we note that the distribution is bimodal (figure \ref{histo_color}). The bi-modality is even stronger when the radio-emitting sub-sample of SFL galaxies is considered (figure \ref{histo_color_rad}). This suggests that the SFL class is actually composed of two subclasses, a bluer one and a redder one with a division at $u-r=2.2$. If we divide the SFL objects by color in this way, we find that the bluer SFL sample has just the right increase with redshift to compensate for the lowering of the SFG fraction (see fig. \ref{dilution_counts_SFL}), while the red SFL sample shows a strong increase, similar to that seen for passive galaxies. This latter group also contains the vast majority of our radio and X-ray AGN with SFL spectra. This suggests that SFL systems with colors redder than $u-r=2.2$ are most likely misclassified emission-line AGN systems with low signal-to-noise spectra and possible pollution from HII regions in their host galaxies (see below). Indeed, we find that the median $u-r$ color ($\sim 2.69$), stellar  mass ($\sim10^{10.8} \msun$), and concentration parameter ($\sim 0.38$) for these systems are all similar to those found for the emission-line AGN population. This population also has the high extinction typical of active galactic nuclei, with a median Balmer decrement calculated from the uncorrected line fluxes of  B=H$\alpha$/H$\beta \sim$ 15.5, between the values found for Seyfert II galaxies (B=11.6) and LINERS (B=20.2) with $u-r>2.2$. In contrast, the blue SFL population has a median Balmer decrement of 8.1, consistent with the idea that these galaxies are extincted star-forming galaxies (for comparison, the median Balmer decrement for the SFG class is B=6.2).  We note, however, that extinction cannot be the primary reason for the misclassification, since the BPT method and its variants consider ratios of adjacent lines and are therefore largely independent of dust extinction.  Furthermore, the line-fluxes have been corrected for dust extinction in \cite{Kauffmann2003}.

As discussed in Ho et al. (2003), there exists another class of active nuclei with LINER-like properties, but which are possibly contaminated by the emission-line light from star-formation in the nuclear region or across the segment of the host galaxy falling into the spectral aperture. These transitional objects are the most likely candidates for the red SFL class. The fact that their fractional abundance increases with redshift (figure \ref{dilution_counts_SFL}) supports the hypothesis that contamination from star-forming regions inside the host pollutes the nuclear light, leading these active galaxies to be misclassified as low signal-to-noise star-forming galaxies: as redshift increases, more light from patches of star-formation falls into the fiber, resulting in their misclassification. 

Indeed, if we consider the trend for composite-systems, which show spectra with both stellar and AGN signatures, we find a steep decrease in their fractional abundance at higher redshifts, as shown in figure \ref{dilution_counts_SFL}. In the same figure we can see also how the decrease in composite systems is nicely balanced by the increase in SFL systems, in agreement with the idea that the red SFL class results from the misclassification of AGN hosted by galaxies with ongoing star-formation. 

As discussed by Ho (2004), transitional objects tend to inhabit bluer systems than pure LINERs and, indeed, we find that the median $u-r$ color for these systems is 2.68, bluer than the value of 2.82 found for the spectroscopically classified emission-line AGN.  Their D4000 and H$\delta$ indicators also suggest a younger stellar population than the hosts of pure AGN. These objects account for 34\% of the total AGN sample after correction, a fraction higher than that found in the Palomar spectroscopic survey of nearby galaxies, suggesting we have a surplus of transitional objects in our sample. Since these cannot be true star-forming galaxies given their distribution with redshift, they must be misclassified AGN in which light from star-forming patches inside the host is increasingly included in the fiber as the host is seen farther away. This would be in agreement with the claim of Kauffmann et al. (2003) that powerful AGN are found in systems with evidence of recent star-formation activity, with young stars residing at larger radii, outside the nuclear region.

We conclude that both  dilution and pollution are  significantly biasing the spectroscopic classification of emission-line galaxies in the SDSS (and other redshift surveys). It is interesting to note that the high percentages of misclassification found suggests that, for spectra with the S/N ratio typical of the SDSS taken with a 3'' fiber, dilution is effective in creating misclassification even for galaxies with sizes much larger than the fiber size, especially in systems lacking strong [OIII] and H$\beta$ lines.

Correcting for this bias is necessary to obtain clean samples of passive galaxies when comparing the properties of the AGN population with a spectroscopically passive sample  (e.g., \cite{Li2006}). Biased star-forming samples given by the introduction of weak-lined AGN would affect the estimate of the local star-formation rates and the study of the distribution of star-forming galaxies across different environments, since AGN tend to populate denser environments than star-forming galaxies.

Furthermore in Paper II we analyze the evolution of active galaxies in the look-back time of the SDSS. In order to compare the properties of nuclei at different redshifts it is necessary to  correct for the spectroscopic  misclassification as a function of redshift. This correction will be also needed in order to compare the \emph{physical} properties of radio vs. spectroscopic active nuclei, the topic of Paper III.

\section{A solution to the problem of dilution}
\label{solution}

In order to select a fair sample of AGN in the local universe we need to find the absorption-line galaxies with low signal-to-noise lines that might have been misclassified because of dilution. To this end we start from the idea that a successful re-classification of such systems must ensure constancy of the fractions of galaxies in each class with redshift.

\begin{figure}
\begin{center}
\includegraphics[scale=0.6,angle=90]{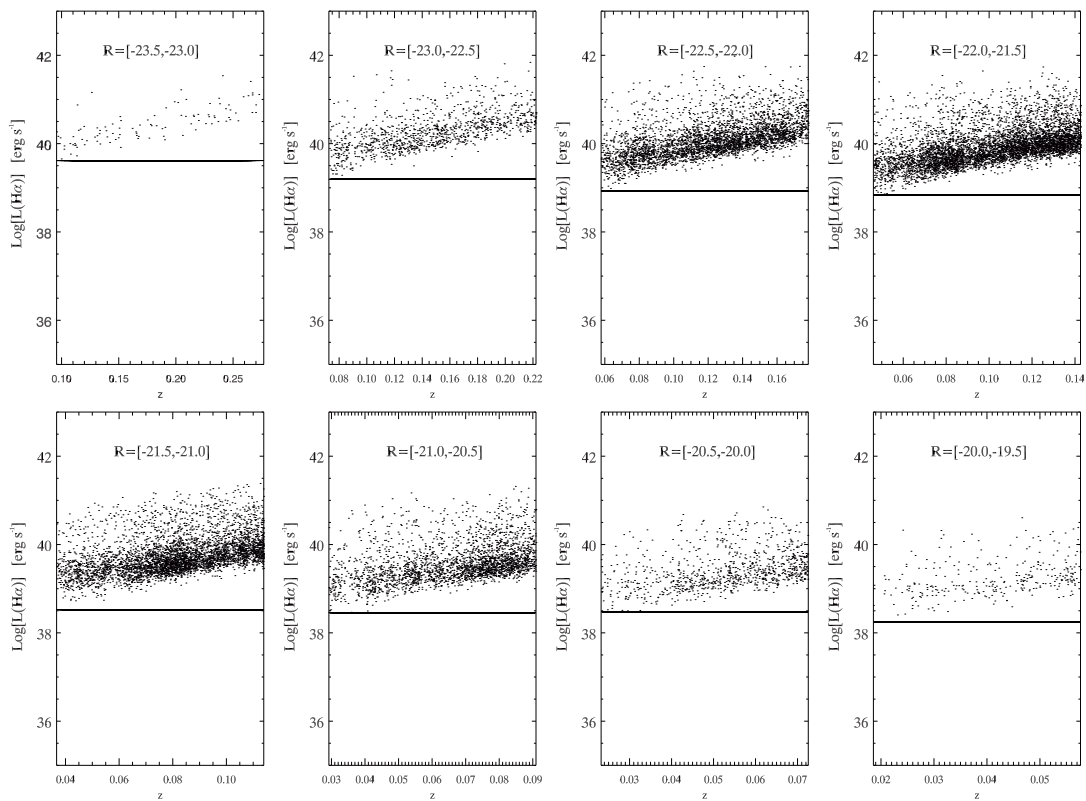}
\end{center}
\caption{\emph{For a chain of eight volume-limited samples, we show the trend of H$\alpha$ luminosity with redshift for spectrally classified AGN. A rising lower cutoff is evident. Low-luminosity AGN are systematically excluded from the sample with increasing redshift. The solid horizontal lines shows the lowest H$\alpha$ luminosity detected each sub-sample.}}
\label{Lum_ha_z_AGN}
\end{figure}

\begin{figure}
\begin{center}
\includegraphics[scale=0.6,angle=90]{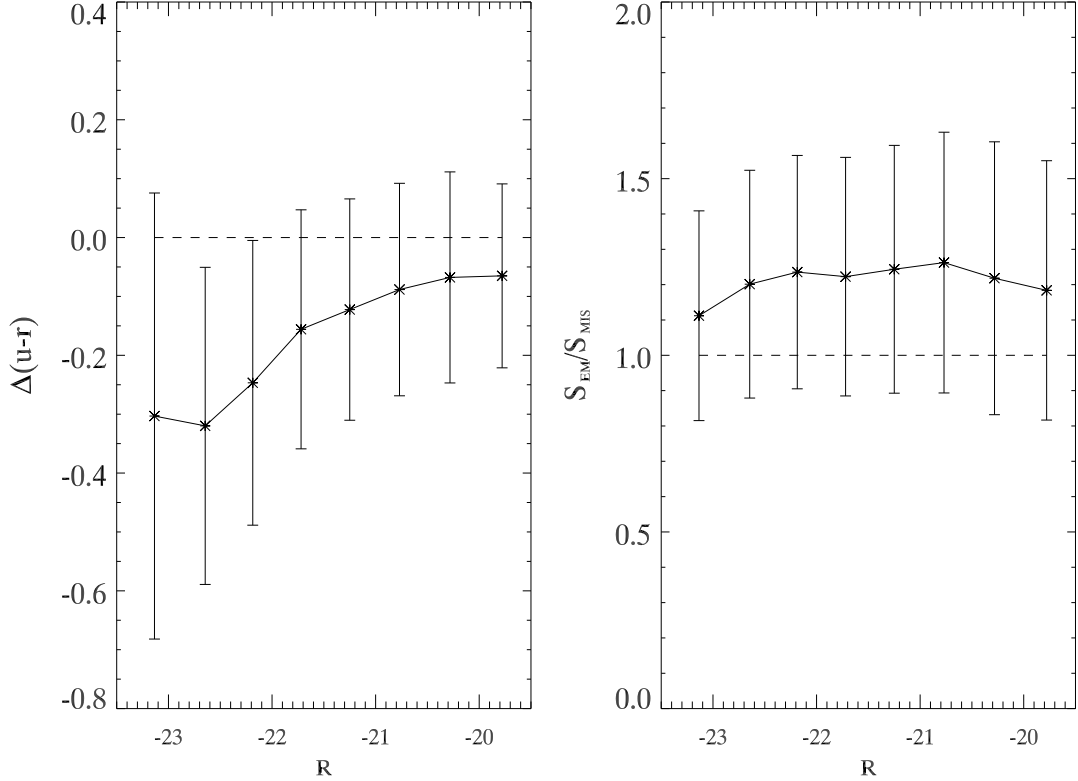}
\end{center}
\caption{\emph{Left: The difference in color between the emission-line AGN population and the misclassified AGN for different magnitude cuts. The misclassified AGN are systematically redder than the spectroscopically classified AGN population. Right: The ratio of the apparent sizes of emission-line AGN to misclassified AGN. The misclassified population is on average 20\% smaller than the emission-line population. Error bars represent the spread in the distributions, calculated as median absolute deviations from the median value. Dashed lines represent the values expected if the two classes were identical.}}
\label{color_size_r}
\end{figure}

Dilution is a redshift-dependent systematic. We have therefore examined the distribution of the luminosity in the H$\alpha$ line strength with varying redshift for emission-line AGN in a chain of eight volume-limited samples covering the absolute magnitude range $-23.5<R<-19.5$ at intervals of 0.5 in absolute magnitude. This ensures that we are looking at optically complete samples of AGN hosts with similar physical properties, since size, color, and mass tend to scale with luminosity and the luminosity bins chosen are narrow.  A rising lower cut-off appears in all luminosity-redshift scatter plots shown in figure  \ref{Lum_ha_z_AGN}. We have also evaluated the median size of the emission-line population and compared it to the median size of the passive population in each volume-limited sample and found that the emission-line galaxies are systematically larger by $\sim$ 20\%  than the ABG systems in all samples (see figure \ref{color_size_r}). This suggests that when hosts of similar luminosity are selected, the more distant ones -- which also appear smaller -- tend to have their lines diluted to the point that, even after continuum subtraction, they are not properly classified. Interestingly, the plots show that the higher luminosity samples, which are selected at higher redshifts, are also the ones that show the largest fraction of missing galaxies, exactly as expected if dilution is at play.

The selection of emission-line systems depends on the signal-to-noise ratio for each line, after continuum subtraction: the more light from the host falling into the aperture, the higher the noise, up to the point that, for severely diluted systems, the signal may not be detectable at all. These systems would be then indistinguishable from truly passive galaxies \cite[]{Moran02}. In order to avoid such misclassification, one would need to require a \emph{decreasing} level in signal-to-noise in the threshold chosen to define an emission-line, instead of choosing a 3$\sigma$ level throughout the whole survey as is generally done. This, of course would come at the price of having less reliable estimates of the fluxes for a substantial fraction of galaxies, but it would avoid significant incompleteness.

Since the careful continuum subtraction performed by Kauffmann et al. (2003) allows a measure of the flux of H$\alpha$ for most absorption-line systems even when the line is fairly unremarkable, in order to correct the data we simply need to find the low signal-to-noise active galaxies misclassified as absorption-line systems.

\begin{figure}
\begin{center}
\includegraphics[scale=0.6,angle=90]{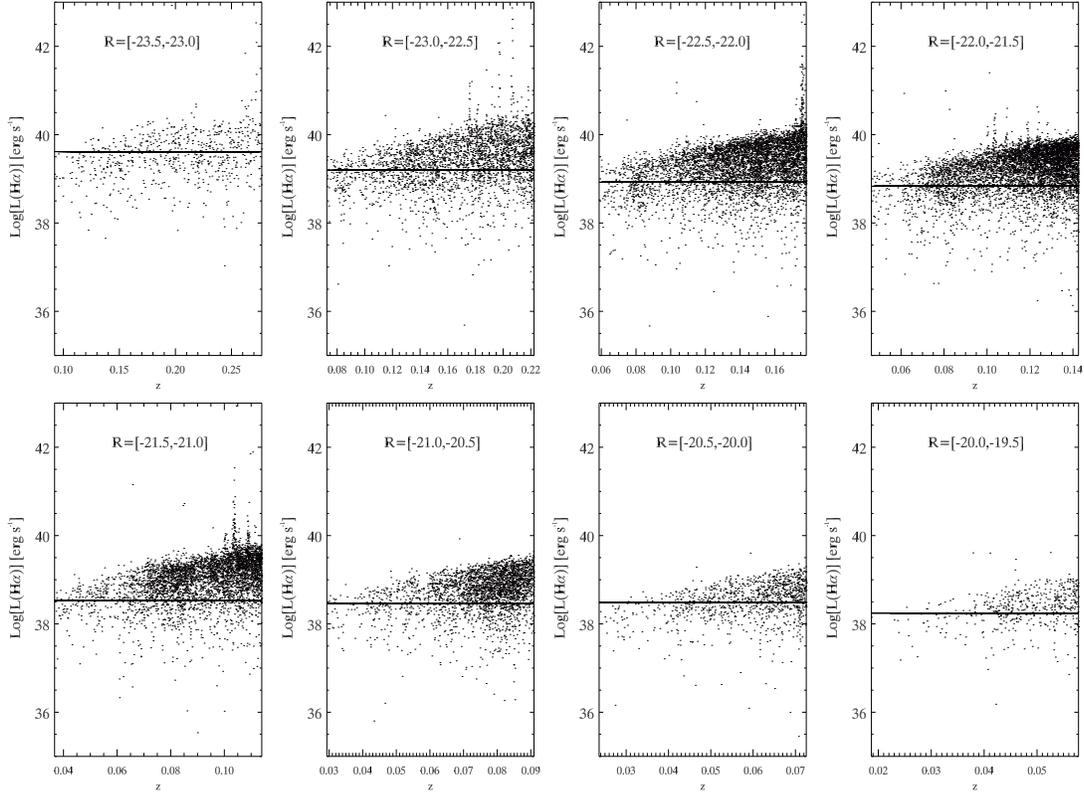}
\end{center}
\caption{\emph{For a chain of eight volume-limited samples we show the trend of H$\alpha$ luminosity with redshift for spectrally classified passive (ABG) galaxies. A systematic excesses of higher luminosity systems is found with increasing redshift. Low-luminosity emission-line systems systematically excluded from the AGN sample are classified as passive galaxies and form the excess of ABG systems with increasing redshift seen in figure 5. The solid line shows the lowest H$\alpha$ luminosity detected in the emission-line AGN sample (from figure 9).}}
\label{Lum_ha_z_ABG}
\end{figure}

In figure \ref{Lum_ha_z_ABG}, we plot the luminosity in H$\alpha$ vs. redshift for ABG galaxies -- the  systems that do not show significant emission-line signatures according to the spectroscopic classification --  and find a population that could compensate for the lack of emission-line AGN in the lower luminosity regimes at higher redshifts. As expected these systems have fairly strong  H$\alpha$  emission, but given the fixed threshold in signal-to-noise adopted in classifying them, they are mistakenly classified as galaxies with unremarkable emission-line signatures.

For each volume-limited sample, the ABG systems with an H$\alpha$ luminosity above the minimum set for the emission-line AGN sample, $L(H\alpha)_{min}$, constitute our best candidates for correcting for the misclassification produced by dilution. However, in order to better select systems with actual emission lines (and not those resulting from poor continuum subtraction or inaccurate reddening corrections),  we have added two more parameters in an attempt to select only the population of galaxies that should be reclassified as AGN.

H$\alpha$ is often one of the strongest emission lines in active galaxies; even if the other lines are diluted, H$\alpha$ might still be detectable. Thus, we selected from among the galaxies with luminosity above $L(H\alpha)_{min}$ those with H$\alpha$ equivalent widths larger than a value $EW(H\alpha)_{min}$. To this cut we added another threshold $\mathcal{R}_{min}$, a lower limit on the fraction of galaxy light falling into the fiber; galaxies with relatively smaller amounts of light falling in the fiber are less likely to be diluted AGN, and are more likely truly passive systems. We varied these two parameters and, for each combination, reclassified the fraction of galaxies satisfying all three requirements. We then chose the combination of parameters which produces the flattest trends in the plots displayed in figures \ref{dilution_counts} and \ref{dilution_radio_AGN}. The final choice is $EW(H\alpha)_{min}=0.31$ and $\mathcal{R}_{min}=0.18$, with the convention that the EW of emission lines is positive. In the right column of the two figures we present the trends after our reclassification is applied. The fraction of AGN and ABG systems are now very similar at each redshift.

\begin{figure}
\begin{center}$
\begin{array}{cc}
\includegraphics[width=2.5in,scale=1.0,angle=0]{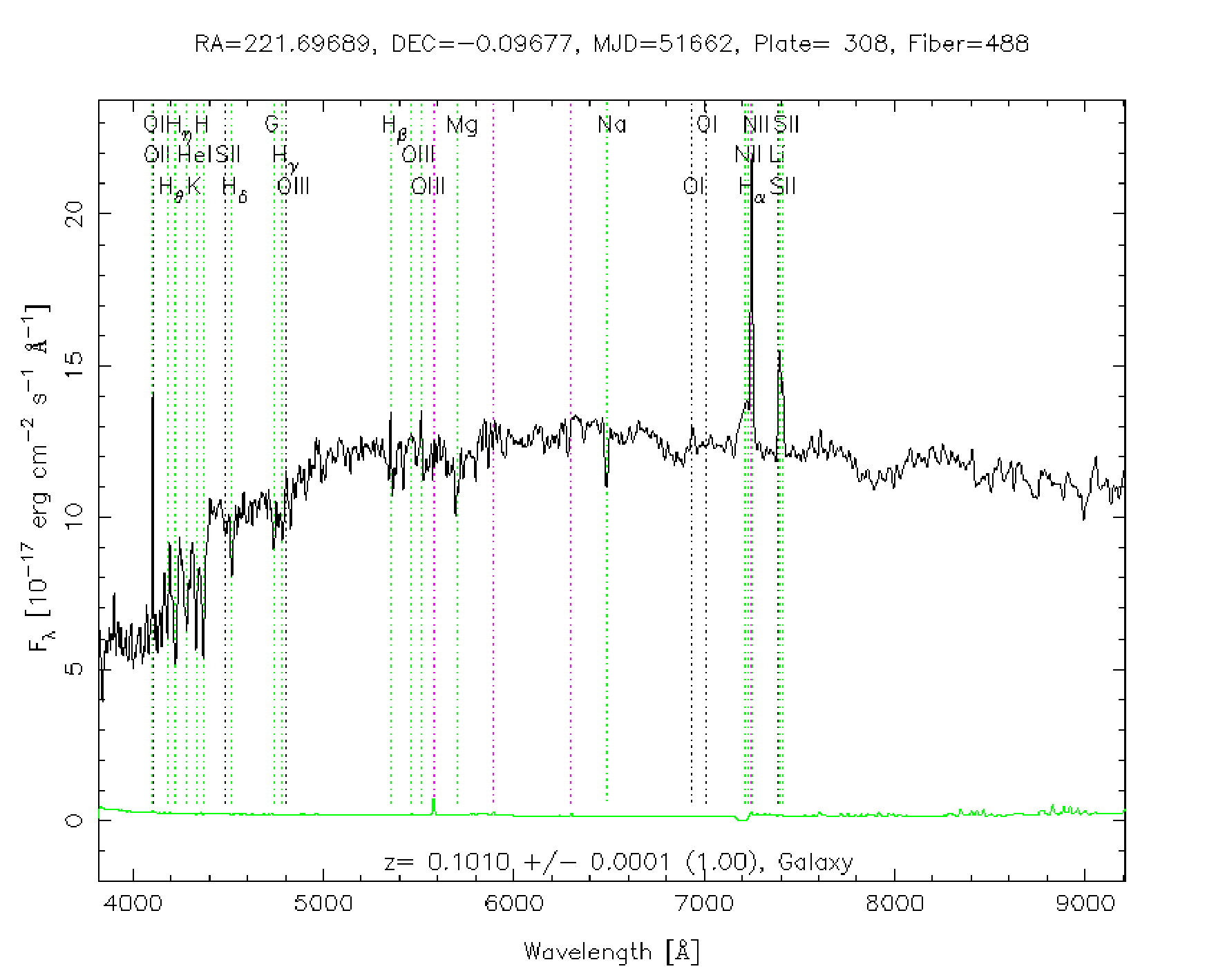}&
\includegraphics[width=2.5in,scale=1.0,angle=0]{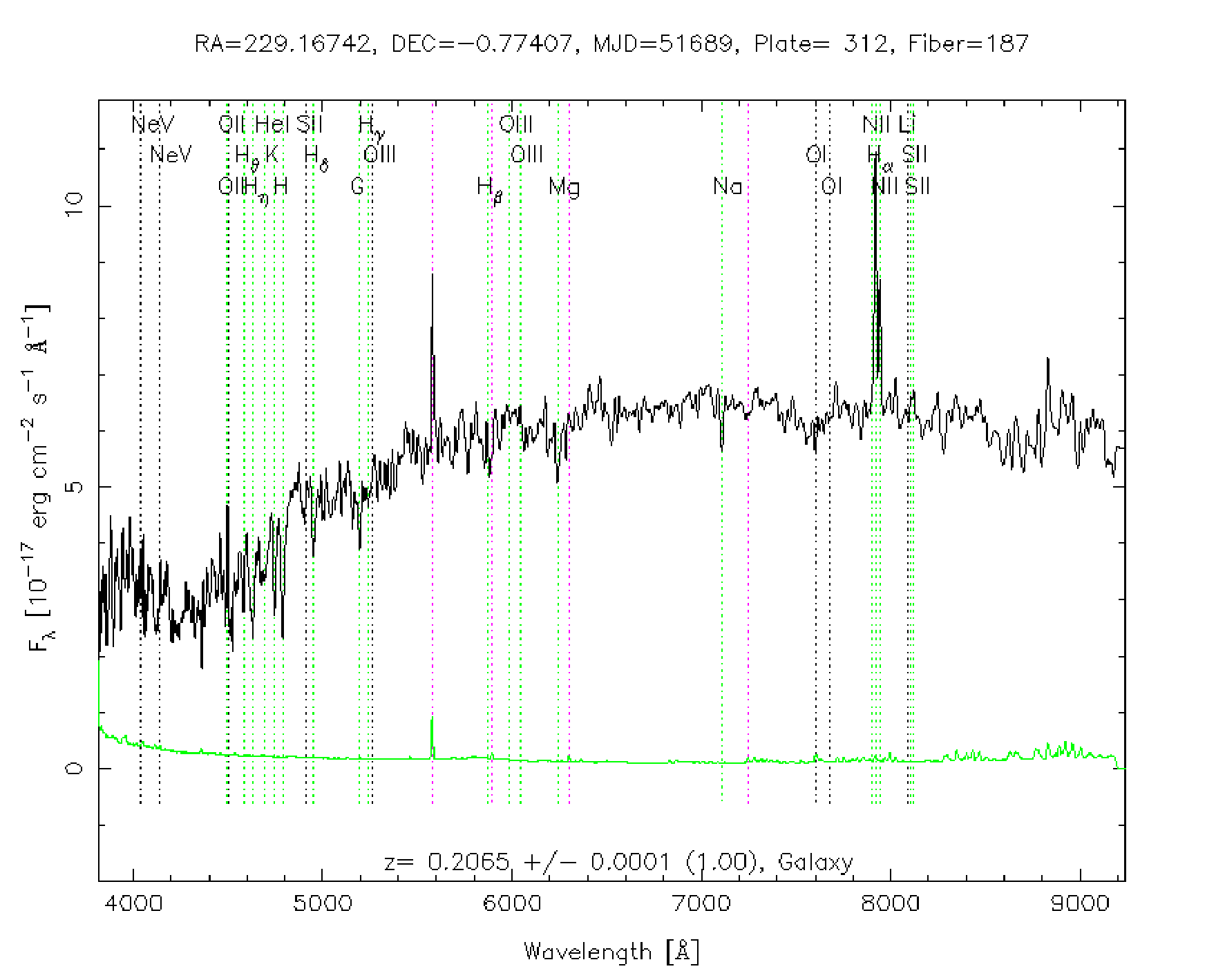} \\ 
\includegraphics[width=2.5in,scale=1.0,angle=0]{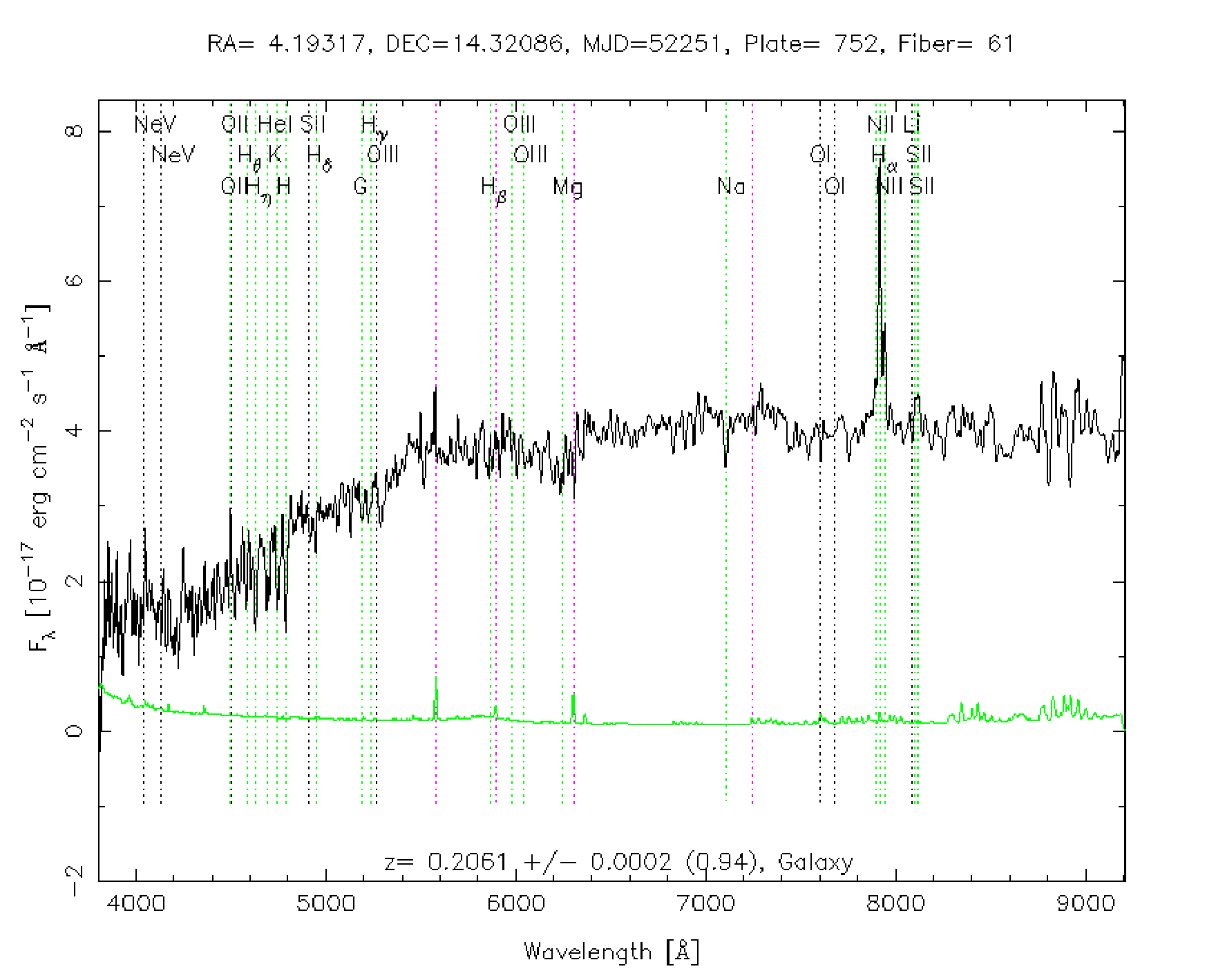} &
\includegraphics[width=2.5in,scale=1.0,angle=0]{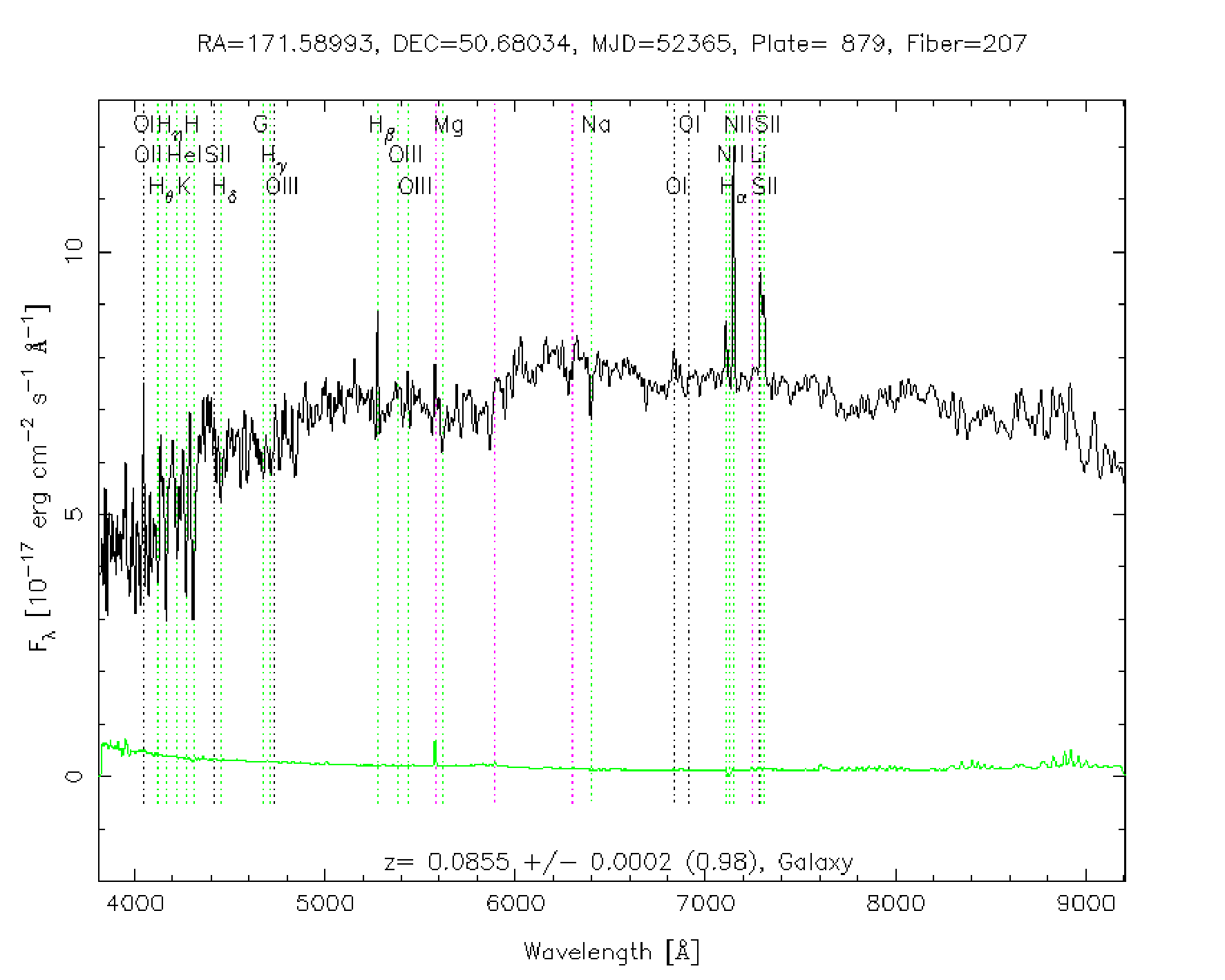}
\end{array}$
\end{center}
\caption{\emph{Spectra for the four galaxies with no [OIII] or H$\beta$ emission, and low signal-to-noise in either H$\alpha$ or [NII] which leaves them unclassified as either AGN or star-forming galaxies. Clockwise from the upper left corner:J144647.24-000548.4, J151640.18-004626.6, J112621.58+504049.1, and  J1001646.36+141915.1}}
\label{no_OIII_spectra}
\end{figure}

   In figure \ref{Lum_ha_z_ABG} we noticed a few systems with suspiciously high H$\alpha$ luminosities. We checked by eye the images and spectra for a sample of ~100 galaxies with $L(H\alpha)>10^{41}$ erg s$^{-1}$. A substantial fraction of these systems appear to have spectra with clear problems affecting the region around H$\alpha$. Some look like passive galaxies and it is likely that their high quoted H$\alpha$ luminosities result from poor continuum subtraction or an incorrect reddening estimate. A more interesting class of objects includes systems in which the H$\alpha$ emission \emph{is} detected, but [OIII], and sometimes the H$\beta$, are not. In figure \ref{no_OIII_spectra} we show spectra for four objects belonging to this class. 

Galaxies with discernible H$\alpha$ and [NII] are also present at lower H$\alpha$ luminosities. An example of an ABG galaxy with L$_{H\alpha}<10^{40}$ erg s$^{-1}$ yet still detectable H$\alpha$ and [NII] lines is shown in figure \ref{spectra_sample}. The spectroscopic criteria for the AGN sample require [NII]6584/H$\alpha>$0.6 and a S/N$>3$ for both lines \cite[]{Brinchmann2004}. Therefore, while some objects have been classified as systems without remarkable emission-lines (ABG), it is not for the lack of [OIII], but the low S/N of the [NII]6584 or H$\alpha$ line.  Of course, not all galaxies in our potential sample of misclassified galaxies showed such visible lines -- indeed, the vast majority did not show emission lines at all.

\section{The spectroscopically unremarkable population}
\label{unremarkable}

Our chain of volume-limited samples includes 104,646 galaxies, $\sim$ 69\% of the total sample. We find that the misclassified galaxies represent roughly 11\% of that total. The number of spectroscopically classified AGN in the whole sample is $\sim$ 23,000. Thus, roughly one-third of all emission-line AGN are misclassified as passive galaxies because of dilution. This quantifies the magnitude of the effect in the SDSS sample. We expect that similar results hold for other surveys unless much smaller apertures are used. This bias does not affect AGN showing all standard emission lines, but preferentially affects systems with weak [OIII] and H$\beta$.

\begin{figure}
\begin{center}
\includegraphics[scale=0.6,angle=90]{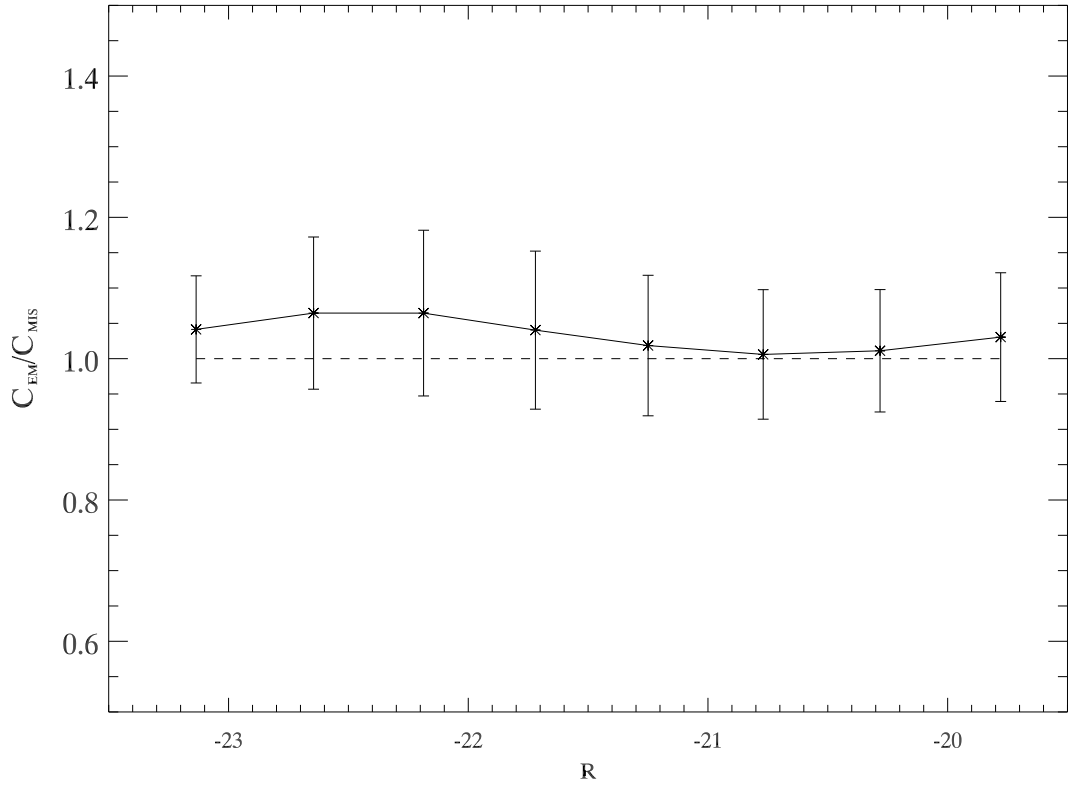}
\end{center}
\caption{\emph{The ratio of the median concentration parameters for classified and misclassified AGN. The ratio is very close to unity for all samples (error bars as in figure 14). }}
\label{conc_flat}
\end{figure}

The average color of the missing AGN population is $u-r=3.00$, somewhat redder than the average color of the emission line population which has $u-r= 2.82$, a value typical of early-type hosts. We note that this is consistent with the fact that, for each volume-limited sample, we find that the emission-line detected population is bluer than the passive population as shown in figure \ref{color_size_r}, and that this effect is more marked in more luminous systems. This suggests that there is an actual physical difference between the two populations, since the photometric color of a galaxy is not influenced by spectroscopic dilution. Redder galaxies tend to have light profiles that are more concentrated and, in principle, it is possible that redder galaxies might therefore suffer greater dilution. However, we show in figure \ref{conc_flat} that the ratio of the
concentration parameter ($C=R_{50}/R_{90}$) for the two populations is
very close to 1, consistent with the notion that, at each magnitude,
we are selecting very similar galaxies. For each magnitude slice we are also then selecting central black holes of roughly the same mass, given the well-established correlation between supermassive black hole mass, absolute R magnitude, and velocity dispersion for spheroidal galaxies \cite[]{Ferrarese2005}. The median stellar mass of the misclassified systems is $10^{11.00}M_{\odot}$, essentially the same as that for the emission-line AGN $10^{10.93}M_{\odot}$. 

Thus, there appears to be an actual physical difference in the nuclear activity between the two populations. One possibility is that lines in redder systems are more extincted by dust, which would help explain why they tend to be more readily omitted from the emission-line sample. Since lines are often not detected in the misclassified group, we cannot compare Balmer decrements to rule out this possibility.  However, we note that heated dust re-radiates in the far-infrared and if these systems had their lines severely dust-extincted, they should show a higher detection rate in the far-infrared than the emission-line population. On the contrary, \emph{at all redshifts}, the fraction of far-infrared-detected sources in the emission-line sample is at least an order of magnitude higher ($\sim$1.5\%) than in the sample of misclassified AGN ($\lesssim 0.1 \%$); this does not support the idea that extinction is the cause of weaker emission lines. In Papers II and III we will explore this issue further, and propose that these systems are starved AGN, in which the accretion rates have become lower, less ionizing radiation is produced and, as a consequence, less intense recombination lines are found. This effect, combined with the higher pollution of their spectra by host light, will cause them to be misclassified when a fixed emission line signal-to-noise ratio is adopted throughout the survey.  A decrease in AGN power for redder systems is consistent with the fact that AGN with high [OIII] luminosities are selected in systems with evidence of recent star-formation (and are therefore bluer), as discussed by \cite{Kauffmann2003}.

We note that the fraction of AGN in early-type systems before any
correction is $\sim$ 23\%. After adding back the misclassified passive
systems, the fraction is $\sim$ 34\%, which means that one-third of the
AGN population in early-type galaxies is missing because of dilution
effects. If we also include the misclassified population of SFL red
galaxies (18\% of the total early-type systems) as mostly AGN on the basis of the foregoing discussion and the
fact that a significant fraction of them shows other indicators of AGN
activity such as strong X-ray or radio emission, we see that missed AGN 
constitute 29\% of the total population of early-type systems: $\sim$11\% 
as a consequence of dilution and 18\% as a result of contamination by
stellar-light from star-forming regions in the host. We thus find the
total fraction of early-type galaxies hosting AGN is 52\%, in excellent
agreement with the Palomar survey \cite[]{Ho2003}. If such a high fraction of early-type
galaxies show signatures of AGN activity, it is reasonable to believe that
most, if not all early-type galaxies undergo an AGN phase in their lives.

The dilution of low-power AGN in early-type galaxies and the misclassification given by star-light pollution will affect estimates for both the luminosity function of AGN from the SDSS and the absolute counts of AGN in the local universe, both of which are crucial in order to compare the high- and low-redshift frequency of activity in galaxies. 

After corrections have been made for the missing population of emission-line AGN, we still find that 6\% of radio and/or X-ray AGN are hosted by truly passive galaxies. These galaxies have fairly large average sizes ($> 7''$) , making dilution an unlikely explanation for the lack of lines. A total of 83\% of these passive objects detected in FIRST show jet-like emission, suggesting that the majority of these lineless systems are associated with strong jet activity, as noted in \cite{Reviglio2006}. These systems are most likely low-excitation radio galaxies \cite[]{Baum1995,Hardcastle2006} which are known to have unremarkable emission lines and often show X-ray emission. For such systems, Evans et al. (2007) suggest that the X-ray emission might be produced by parsec scale jet emission. A fairly large population of galaxies lacking emission lines but with AGN-level x-ray emission has also been noted in clusters by Martini et al. (2002; 2007).  

\begin{figure}
\begin{center}
\includegraphics[scale=0.6,angle=90]{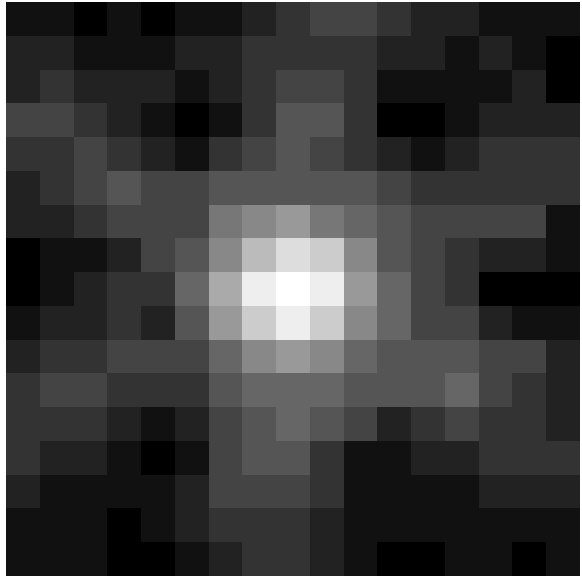}
\end{center}
\caption{\emph{The median radio source obtained by stacking in flux density  the fields of 28,141 passive galaxies without a FIRST detection. The median luminosity is 6.2$\times 10^{27}$ erg s$^{-1}$ Hz$^{-1}$}}
\label{ABG_true_stack}
\end{figure}

\begin{figure}
\begin{center}
\includegraphics[scale=0.6,angle=90]{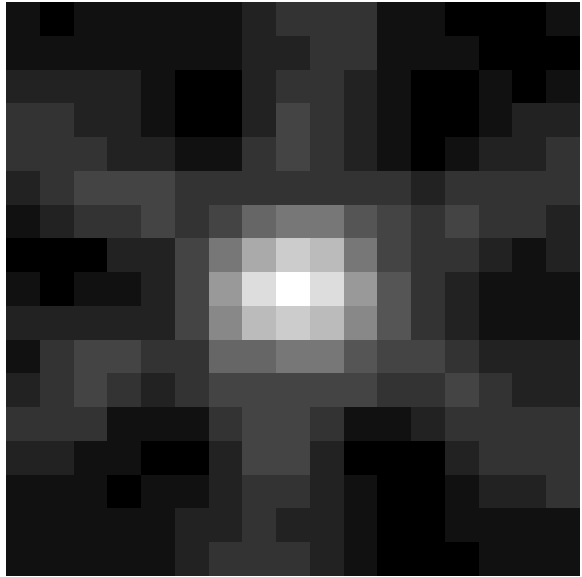}
\end{center}
\caption{\emph{The median radio source obtained by stacking in flux density the fields of 21,454 emission-line AGN without detection in FIRST. The median luminosity is 1.2$\times 10^{28}$ erg s$^{-1}$ Hz$^{-1}$}}
\label{AGN_stack}
\end{figure}

In these systems, which tend to inhabit denser environments, the lines appear to be truly missing. Since the radio survey only detects the strongest radio sources given its sensitivity threshold, it is possible that a large fraction of passive systems have low-level radio AGN activity without associated emission lines.  In figure \ref{ABG_true_stack} we show that nuclear radio activity in this population of spectroscopically unremarkable galaxy hosts is not an exception: by stacking FIRST radio images of the undetected population at $z<0.2$ using the procedure described in \cite{White2007}, we are able to measure a median flux density of $32\mu$Jy and a median radio luminosity of $6.2\times 10^{27}$ erg s$^{-1}$ Hz$^{-1}$ for this population, two orders of magnitude fainter than detected point-like radio population which has median radio luminosity of $7.6\times10^{29}$erg s$^{-1}$ Hz$^{-1}$. The undetected sample's median redshift is 0.118, close to the median redshift (z=0.115) of the emission-line AGN subsample undetected in FIRST, suggesting that most of these radio-emitting nuclei are not detected in FIRST because of lower instrinsic luminosities and \emph{not} because they are located at greater distances. Hidden, low-luminosity radio AGN in spectroscopically unremarkable nuclei of massive red galaxies may therefore be the norm. For comparison, in figure \ref{AGN_stack} we show the stacked image for the population of emission-line AGN not detected in the radio; the median flux density is $86\mu$Jy, yielding a median luminosity $1.2\times 10^{28}$ erg s$^{-1}$ Hz$^{-1}$ for the median redshift of 0.094, within a factor of two of the passive galaxy non-detections and $\sim$ 60 times less luminous than the detected AGN. This shows that there also exists a population of emission line systems with radio emission undetected in FIRST because of low but distinctly non-zero  radio luminosities.

The fact that lineless radio and X-ray AGN are not accounted for in the luminosity functions of AGN determined spectroscopically (\cite{Hao2005}) is another bias that needs to be taken into consideration. In particular, since these galaxies are likely to host the most massive black holes according to standard bulge-black-hole-mass relations, it is important to understand why this population does not show significant lines. We will address this problem in Papers II and III.

\section{Activity and local density}
\label{environment}

\begin{figure}
\begin{center}
\includegraphics[scale=0.6,angle=90]{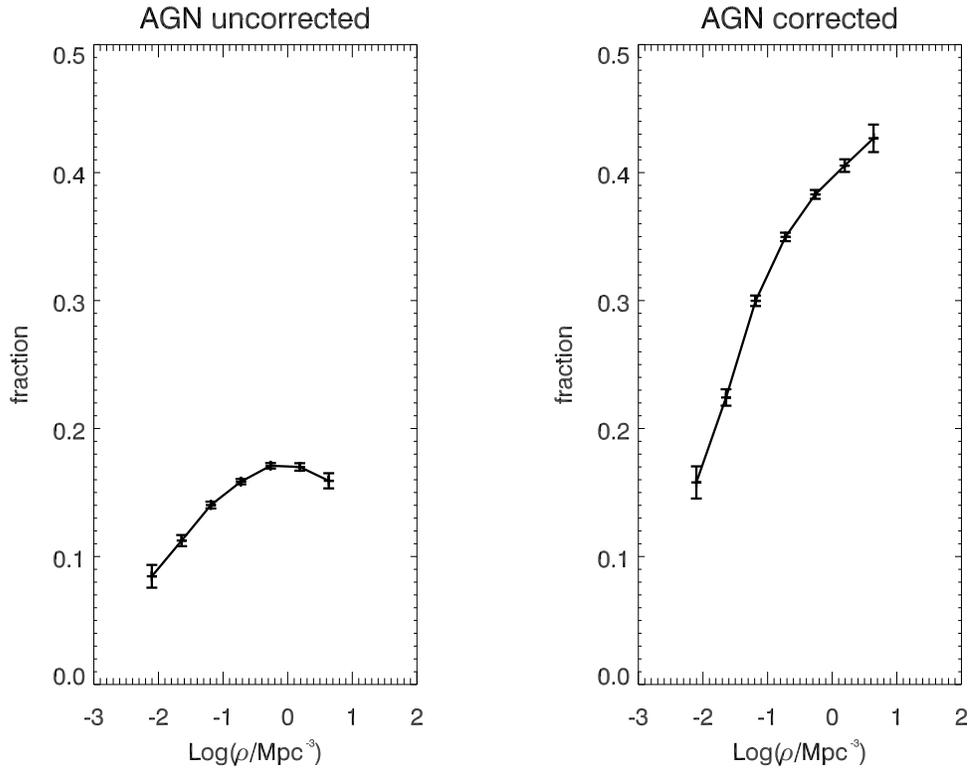}
\end{center}
\caption{\emph{The fraction of spectroscopically classified AGN as a function of local galaxy density before (left) and after (right) correction for misclassification.}}
\label{compare_density}
\end{figure}

Galaxies suffering from dilution are found in all environments, with a slight excess in denser environments compared to the emission-line AGN. In figure \ref{compare_density} we show the trend for the fractional abundance of all spectroscopic AGN with z$<$0.2 as a function of local galaxy density before and after correction. Inclusion of the optically unremarkable class of AGN makes the trend rise steeply. In particular, after correction, the fraction continues to rise into the densest environments. This trend is qualitatively similar to that found using radio emission to select AGN. Since AGN are hosted primarily by early-type galaxies, and the fraction of these increases steeply with density, this trend suggests that a similar fraction of early-type galaxies powers AGN in all environments.  This is in striking  disagreement with previous work using spectroscopic samples by Carter et al. (2001) and Miller et al. (2005).  A rising trend can be seen in the SDSS sample even before correction: the careful subtraction of stellar light by Kauffmann et al. (2003) reveals the weak emission-line population in denser environments. After correction for dilution, this trend becomes much stronger and does not turn over in denser environments. 

\begin{figure}
\begin{center}
\includegraphics[scale=0.6,angle=90]{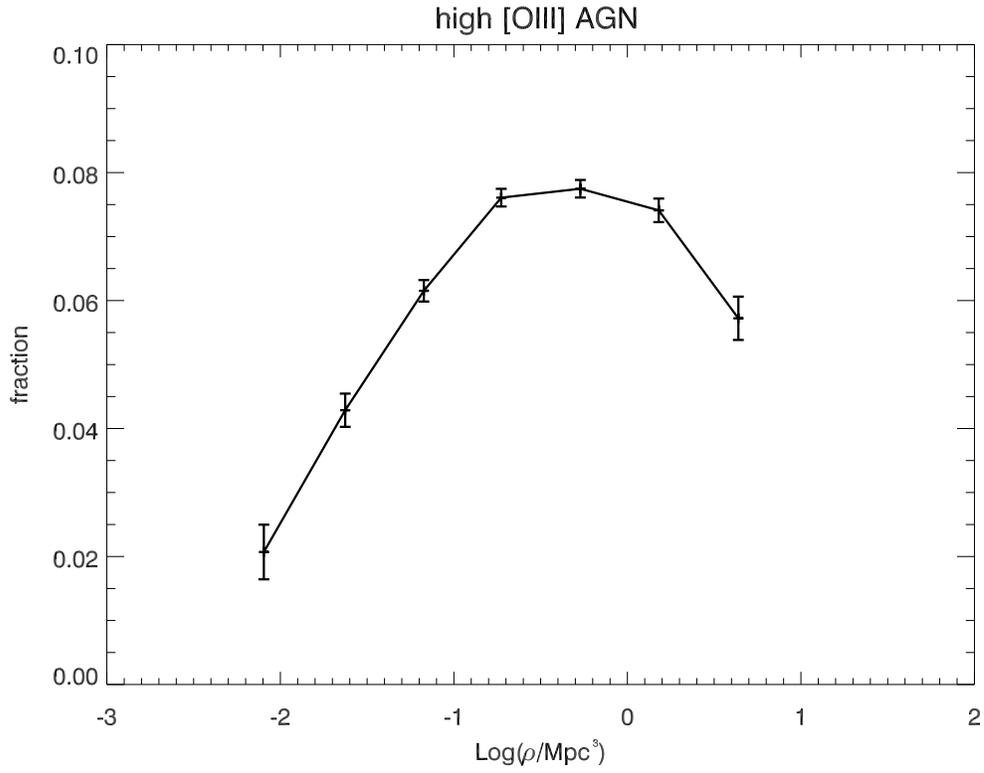}
\end{center}
\caption{\emph{The fractional abundance of strong high $L_{[OIII]}>10^6 L_{\odot}$-luminosity AGN. As discussed by \cite{Kauffmann2004}, strong AGN are less frequent in denser environments.  }}
\label{highOIII}
\end{figure}

This steep increase in the fractional abundance of AGN is not in conflict with the claim by \cite{Kauffmann2004} that strong AGN activity decreases with increasing environmental density. As we show in figure \ref{highOIII}, if we select AGN with [OIII] luminosities above $10^6 L_{\odot}$, we indeed find a decrease in fractional abundance in denser environments. In particular, our density estimator allows us to explore the distribution at all densities, since we are able to associate an environmental density with each galaxy, and thus have a much larger statistical sample than the one used by \cite{Kauffmann2004}. The distribution is suppressed at low densities, peaks at intermediate densities ($\sim$ 0.1 galaxies/Mpc$^{-3})$ and then decreases for higher densities.

Combining this with our previous results, it appears clear that AGN in denser environments have \emph{weaker} emission-line activity, but are \emph{more} frequent. We will further explore this fact in a subsequent paper, where we analyze the interplay between physical properties and environment for different spectroscopic classes of objects.

\section{Conclusion}
\label{conclusion}

In this first installment of our study of active galaxies in the local universe, we have explored the incidence of spectroscopically unremarkable galaxies in the Sloan Digital Sky Survey. These galaxies may not exhibit the expected emission lines either because of systematics affecting the sample, or because of different physical processes at play in their nuclei. Separating these two effects is critical for understanding activity in galaxies.

We have developed a method to evaluate the effect of dilution on spectroscopically selected active galaxies and then have used it to show that samples classified and selected solely on the basis of optical spectroscopy are biased in that they exclude a larger number of low signal-to-noise emission-line AGN with increasing distance. Even when sophisticated templates for stellar light subtraction are used to extract the emission-line spectrum coming from the nucleus, standard diagnostics of activity become more and more biased as systems at greater distances are included, since dilution from stellar light and pollution from star-forming regions suppress or modify the emission-line signature of the AGN. This affects the selection of clean control samples of passive galaxies and clean samples of star-forming ones and might bias the conclusions of a number of studies where uncorrected samples are used.  Early-type systems which are generally selected at higher redshift in magnitude-limited surveys, being the more luminous galaxies, are the objects that suffer most from this effect, with a fraction of undetected spectroscopic AGN as high as 29\%. This effect is not produced by dust obscuration, which in itself might further bias the evaluation of AGN demographics in the local universe. When the misclassified objects are added, we find that up to $\sim$ 52\% of early-type galaxies show evidence of AGN activity.

We have shown that, for a fixed host luminosity, redder hosts tend to have their lines more diluted. This suggests a physical difference in the AGN  between bluer and redder systems, in agreement with the claim that AGN with stronger [OIII] emission are associated with systems showing evidence of recent star-formation.We have argued that this is not an effect of dust-extinction, since the far-IR detection statistics for the two populations do not support this explanation. Nor is this an effect of different black hole masses or light concentration in these systems, since we have drawn our comparisons from volume-limited samples spanning narrow ranges in R-band luminosities (and therefore black hole masses and light concentration parameter), and have shown that the effect is still present when very narrow ranges in host luminosities are selected, a fact that  suggests a link between the age of the stellar population and the strength of the spectroscopic lines. In Paper II we show  evidence of co-evolution of host and nuclear properties, in agreement with this interpretation.

We have explored the distribution of environments in which AGN are found.
While we confirm that the fraction of AGN with strong [OIII] emission decreases at high densities, we show that including low-luminosity AGN and the class of spectroscopically undetected AGN modifies the trend of the fractional AGN abundance with environmental density: AGN are \emph{more} frequent in denser environments, but their spectroscopic signatures become \emph{less} remarkable. The increasing  fractional trend is similar to that found for radio AGN by \cite{Best2005} and  \cite{Reviglio2006}, and is most likely linked to the well-known, although poorly understood, density-morphology correlation, since AGN are typically found in association with early-type galaxies, a fact that we will explore fully in a subsequent paper. 

After accounting for the population of diluted hosts, we find that a fairly large population of radio and X-ray AGN with no spectral signatures is still present, and we conclude that this is indicative of a true difference in host properties. Given the relatively low sensitivities of radio and X-ray surveys, AGN may be present in up to 80\% of the spectroscopically unremarkable population. Indeed, we have shown that, when this population of truly unremarkable galaxies is stacked, nuclear radio emission is detected. This suggests that the vast majority of early-type galaxies undergo an AGN phase in their lives. Comparing the spectroscopic and radio populations of AGN and exploring the physical reasons behind their differences, and the evolution of their differences over time is the subject of Papers II and III.

\section{Acknowledgments}

We wish to thank Jacqueline van Gorkom, Bob Becker, David Schiminovich,  Ben Johnson and an anonymous referee  for helpful discussions and suggestions. This work has made extensive use of the Vizier Online Service. PMR was supported by the National Science Foundation under grant AST-06-07643.

Funding for Sloan Digital Sky Survey project has been provided  by the Alfred P. Sloan Foundation, the Participating Institutions, the National Aeronautics and Space Administration, the National Science Foundation, the U.S. Department of Energy, the Japanese Monbukagakusho, and the Max Planck Society. The SDSS website is http://www.sdss.org .

The SDSS is managed by the Astrophysical Research Consortium (ARC) for the Participating Institutions (The University of Chicago, Fermilab, the Institute for Advanced Study, the Japan Participation Group, The Johns Hopkins University, the Korean Scientist Group, Los Alamos National Laboratory, the Max-Planck-Institute for Astronomy, the Max-Planck-Institute for Astrophysics, New Mexico State University, University of Pittsburgh, University of Portsmouth, Princeton University, the United States Naval Observatory, and the University of Washington).

\clearpage

\clearpage

\clearpage

\clearpage

\clearpage

\clearpage

\clearpage

\clearpage

\clearpage
\clearpage

\clearpage

\clearpage

\clearpage

\clearpage

\clearpage

\clearpage

\clearpage

\clearpage

\clearpage

\clearpage

\end{document}